\documentclass[3p]{elsarticle}
\pdfoutput=1
\usepackage{lineno}
\PassOptionsToPackage{hyphens}{url}\usepackage[colorlinks=true]{hyperref}

\usepackage{amsfonts}
\usepackage{amsmath}
\usepackage{amssymb}
\usepackage{multirow}
\usepackage[hang, perpage, flushmargin]{footmisc}

\usepackage{threeparttable}

\biboptions{authoryear}

\begin{document}

\begin{frontmatter}

\title{ROOD-MRI: Benchmarking the robustness of deep learning segmentation models to out-of-distribution and corrupted data in MRI}

\author[1,2,3]{Lyndon Boone}
\ead{lyndon.boone@mail.utoronto.ca}

\author[2,3,4]{Mahdi Biparva}

\author[2,3,4]{Parisa Mojiri Forooshani}

\author[2,3,4]{Joel Ramirez}

\author[2,4,5]{Mario Masellis}

\author[6,7]{Robert Bartha}

\author[2,3,8]{Sean Symons}

\author[1,9]{Stephen Strother}

\author[2,4,5]{Sandra E. Black}

\author[2,3,8]{Chris Heyn}

\author[1,3]{Anne L. Martel}

\author[2,4,5]{Richard H. Swartz}

\author[1,2,3,4]{Maged Goubran\corref{correspondingauthor}}
\cortext[correspondingauthor]{Corresponding author}
\ead{maged.goubran@utoronto.ca}

\address[1]{Department of Medical Biophysics, University of Toronto, Toronto, Canada}
\address[2]{Hurvitz Brain Sciences Research Program, Sunnybrook Research Institute, Toronto, Canada}
\address[3]{Physical Sciences, Sunnybrook Research Institute, Toronto, Canada}
\address[4]{Canadian Partnership for Stroke Recovery, Heart and Stroke Foundation, Toronto, Canada}
\address[5]{Department of Medicine, University of Toronto, Toronto, Canada}
\address[6]{Department of Medical Biophysics, Western University, London, Canada}
\address[7]{Robarts Research Institute, Western University, London, Canada}
\address[8]{Department of Medical Imaging, University of Toronto, Toronto, Canada}
\address[9]{Rotman Research Institute, Baycrest, Toronto, Canada}

\begin{abstract}
Deep artificial neural networks (DNNs) have moved to the forefront of medical image analysis due to their success in classification, segmentation, and detection challenges. A principal challenge in large-scale deployment of DNNs in neuroimage analysis is the potential for shifts in signal-to-noise ratio, contrast, resolution, and presence of artifacts from site to site due to variances in scanners and acquisition protocols. DNNs are famously susceptible to these \textit{distribution shifts} in computer vision. Currently, there are no benchmarking platforms or frameworks to assess the robustness of new and existing models to specific distribution shifts in MRI, and accessible multi-site benchmarking datasets are still scarce or task-specific. To address these limitations, we propose ROOD-MRI: a novel platform for benchmarking the \underline{R}obustness of DNNs to \underline{O}ut-\underline{O}f-\underline{D}istribution (OOD) data, corruptions, and artifacts in \underline{MRI}. This flexible platform provides modules for generating benchmarking datasets using transforms that model distribution shifts in MRI, implementations of newly derived benchmarking metrics for image segmentation, and examples for using the methodology with new models and tasks. We apply our methodology to hippocampus, ventricle, and white matter hyperintensity segmentation in several large studies, providing the hippocampus dataset as a publicly available benchmark. By evaluating modern DNNs on these datasets, we demonstrate that they are highly susceptible to distribution shifts and corruptions in MRI. We show that while data augmentation strategies can substantially improve robustness to OOD data for anatomical segmentation tasks, modern DNNs using augmentation still lack robustness in more challenging lesion-based segmentation tasks. We finally benchmark U-Nets and transformer-based models, finding consistent differences in robustness to particular classes of transforms across architectures. The presented open-source platform enables generating new benchmarking datasets and comparing across models to study model design that results in improved robustness to OOD data and corruptions in MRI.
\end{abstract}


\end{frontmatter}


\section{Introduction}
\label{sec:intro}

Segmentation and volumetry of medical images is a fundamental analysis step in research and clinical settings, as the quantification and localization of anatomical and pathological structures are used in diagnosis, prognosis, and treatment planning for various conditions \citep{giorgio2013clinical, bauer2013survey}. Recently, deep artificial neural networks (DNNs) have moved to the forefront of medical image analysis, achieving state-of-the-art results in segmentation, classification, and detection tasks relating to diverse anatomical regions and imaging modalities \citep{litjens2017survey, yi2019generative}. Across a broad spectrum of tasks, these deep learning methods frequently outperform their classical machine learning counterparts by considerable margins and achieve or surpass human performance in some applications \citep{he2015delving, kooi2017large, esteva2017dermatologist}.

While DNNs continue to achieve impressive results in computer vision, they are known to have various vulnerabilities making them less robust than the human visual system \citep{geirhos2018generalisation}. \citet{szegedy2013intriguing} discovered that DNNs are susceptible to \textit{adversarial examples}: images corrupted with imperceptible noise that fool the network into making false predictions with high confidence. While adversarial examples have attracted considerable interest in the medical imaging research community \citep{paschali2018generalizability, li2020anatomical, ma2021understanding, xu2021towards, daza2021towards}, critics have questioned their relevance in the clinical and research settings since the noise in these images must be carefully designed by an intelligent system. Meanwhile, more prevalent distortions affecting natural image quality, such as random noise, blurring, and contrast alterations, have been thoroughly demonstrated to degrade DNN performance in the broader computer vision literature \citep{dodge2016understanding, geirhos2018generalisation}, yet are mostly understudied in medical imaging. For a model trained on clean data, these distortions seen at test time constitute an example of \textit{distribution (or domain) shift} -- broadly defined as a discrepancy between training and test data characteristics \citep{quinonero2008dataset}. Test data constituting this shift are called \textit{out-of-distribution (OOD) data}.

Domain shifts are ubiquitous in multi-site and multi-scanner MR image analysis, where models are often trained on labeled data from one or a few sites before being tasked with making predictions on data from extrinsic sites \citep{karani2021test}. Unlike CT, MR voxel intensities are not standardized to a scale corresponding to a physical material property, making their distributions highly variable across scanners. Field strength, fine control over several acquisition parameters, and MR pulse sequence design can greatly influence the relative contrast between tissue types and image resolution. Differences between scanner setups and hardware can manifest as variations in signal-to-noise ratio (SNR) and intensity non-uniformity within an image \citep{cardenas2008noise, sled1998understanding}. Moreover, MRI is associated with unique imaging corruptions and artifacts that arise from the acquisition process (e.g., k-space motion artifacts) \citep{zhuo2006mr}. Often, scans with mild artifacts are removed from training and validation datasets, making them OOD at test time in scenarios where they must be analyzed (e.g., when a retest scan is unavailable). While many approaches have attempted to correct for artifacts and corruptions \citep{tustison2010n4itk, godenschweger2016motion, usman2020retrospective}, adapt the model to the unseen domain \citep{karani2021test, guan2021domain}, or improve model robustness through exposure to simulated data during training \citep{zhang2020generalizing}, the field still lacks a basic understanding of how specific distribution shifts and corruptions in MRI can affect DNN segmentation performance.

Notably, there is currently no benchmarking platform and methodology in the field to assess the robustness of new and existing models to specific distribution shifts. In natural image classification, \citet{hendrycks2019benchmarking} created robustness benchmarking datasets from ImageNet \citep{deng2009imagenet} and CIFAR-10\footnote{\url{https://www.cs.toronto.edu/~kriz/cifar.html}} by applying 15 unique corruptions to images at five severity levels each, including various types of noise, contrast alterations, and blurring/pixelation. This benchmarking approach has since been extended to object detection tasks \citep{michaelis2019benchmarking} and semantic segmentation \citep{kamann2020benchmarking}. However, similar studies with benchmarking datasets in medical imaging are sparse. While \citet{campello2021multi} provided a multi-site and multi-scanner challenge dataset intended to benchmark models on their generalizability to data from extrinsic sites and scanners relative to the training set, they did not investigate which characteristics of the extrinsic data or distribution shifts led to performance degradation. Moreover, the scope of their dataset was limited to cardiac MRI. Understanding DNN sensitivities to particular types of OOD data and corruptions could be of great value for designing more robust networks and in contexts where the specific distribution shifts between sites or datasets are known.

To address these issues, we propose ROOD-MRI: a novel benchmarking platform for evaluating the \underline{R}obustness of deep networks to \underline{OOD} data, corruptions, and artifacts in \underline{MRI}. To the best of our knowledge, our work is the first to present a platform tackling DNN robustness to specific distribution shifts and corruptions in MRI and neuroimaging segmentation. Our contributions are summarized as follows:
\begin{itemize}
    \item We develop a methodology for benchmarking DNNs based on robustness to distribution shifts and corruptions in MR neuroimaging segmentation tasks. Our platform (\url{https://github.com/AICONSlab/roodmri}) includes methods for simulating distribution shifts in existing datasets at varying severity levels using imaging transforms (Section \ref{sec:transforms}). 
    \item We generate benchmarking datasets for hippocampus, ventricle, and white matter hyperintensity (WMH) segmentation (Section \ref{sec:tasks_datasets}), providing the hippocampus dataset as a publicly available benchmark. We publish modules for generating new benchmarking datasets that are not confined to specific segmentation tasks or sequences.
    \item We propose novel overlap- and distance-based robustness metrics suited to medical image segmentation to compare new or existing models, accounting for degradation in the central tendency and variance of model predictions when faced with OOD data (Section \ref{sec:metrics}). 
    \item By evaluating U-Net and transformer-based models on these three datasets, we demonstrate that modern DNNs are highly susceptible to distribution shifts and corruptions in MRI. Furthermore, we find consistent differences in robustness to particular classes of transforms between the two architectures, offering insights into fully convolutional vs. transformer-based processing (sections \ref{sec:baseline_sensitivity} and \ref{sec:res:architectures}).
    \item We perform systematic experiments to quantify improvements in OOD robustness associated with data augmentation during training. We show that while data augmentation strategies can substantially improve robustness to OOD data for anatomical segmentation tasks, modern DNNs using augmentation still lack robustness in more challenging lesion-based segmentation tasks such as WMH segmentation (Section \ref{sec:augmentation_results}). Furthermore, we compare the OOD robustness of patch-based and whole-image models, demonstrating that patch size and spatial context play a role in robustness to a particular class of transforms (Section \ref{sec:patch_results}).
\end{itemize}

\section{Methods}
\label{sec:methods}

\begin{figure}[t!]
\centering
\includegraphics[width=\textwidth]{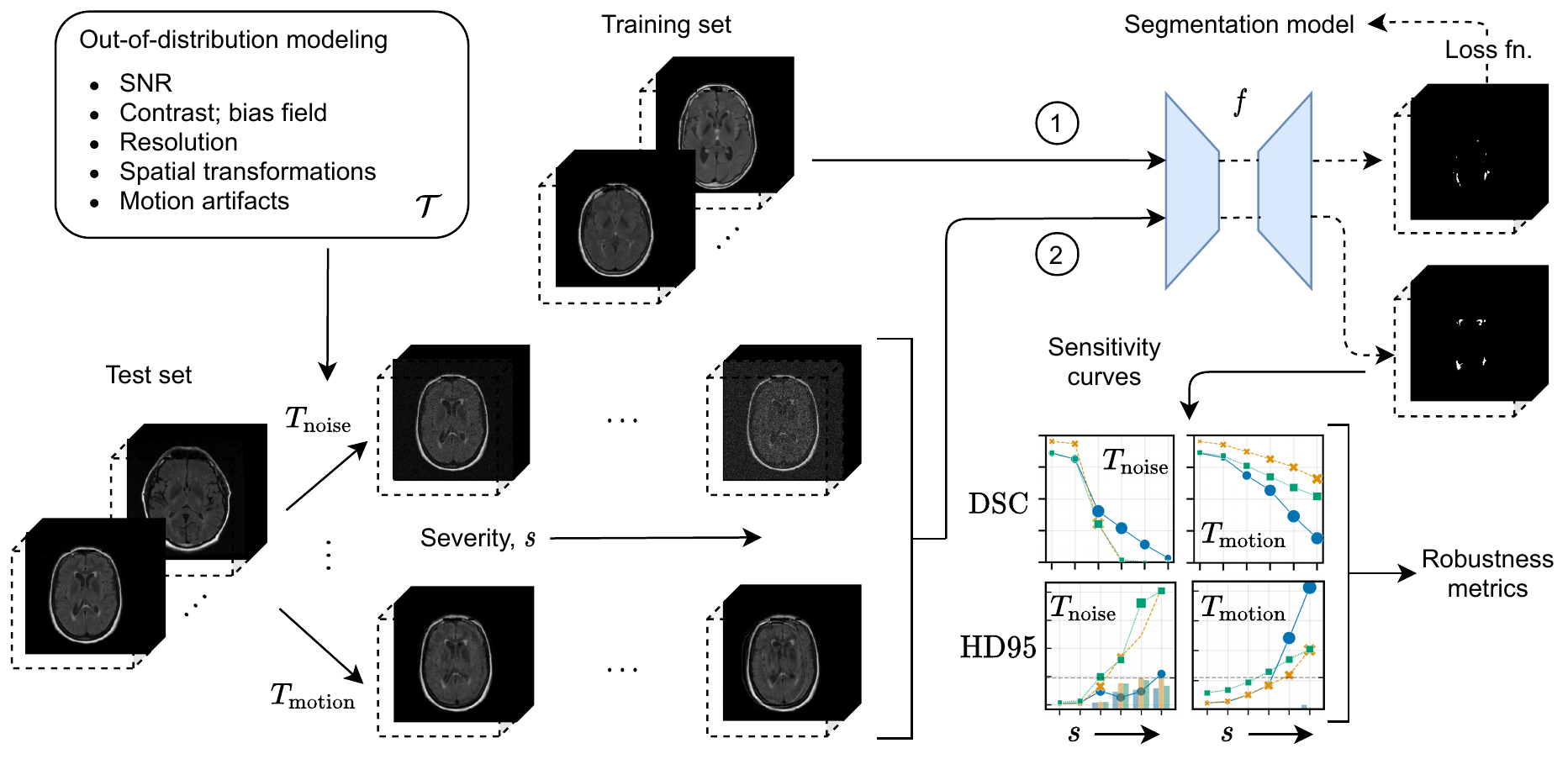}
\caption{Methodology overview for benchmarking robustness to OOD data and corruptions. A segmentation model $f$ is first trained on the clean training set (path 1). The test set is expanded by applying OOD-modeling transforms $T_i \in \mathcal{T}$ to each sample, with defined severity levels $s$ for each transform (path 2). The trained segmentation model $f$ generates predictions on each transformed test sample, yielding robustness metrics used for benchmarking. SNR: signal-to-noise ratio; DSC: Dice similarity coefficient; HD95: modified (95th-percentile) Hausdorff distance.}
\label{fig:methodology_overview}
\end{figure}

\subsection{Transforms}
\label{sec:transforms}

\begin{figure}[t!]
\centering
\includegraphics[width=\textwidth]{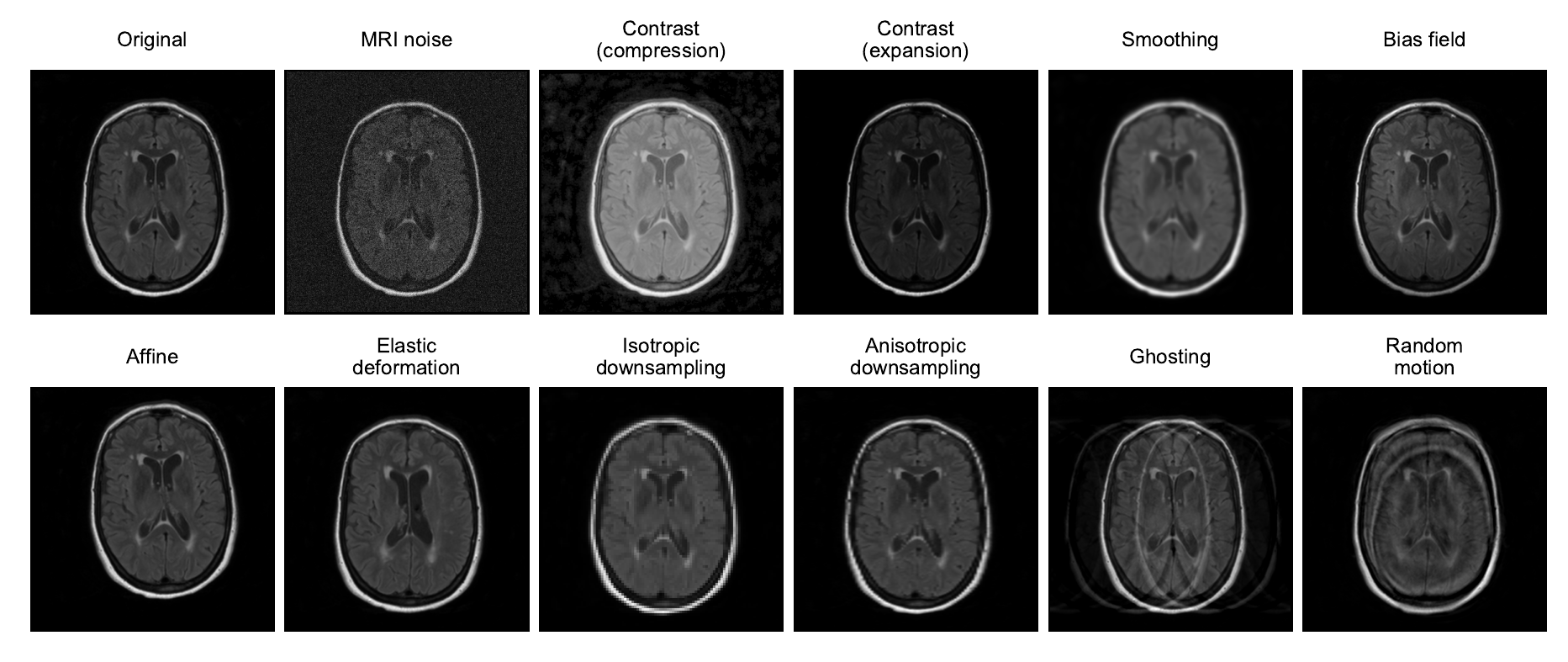}
\caption{Transforms used to model OOD data, corruptions, and artifacts in MRI. In this figure, transforms are applied to a FLAIR image for a patient with periventricular WMH lesions (original image prior to transformations shown in top left).}
\label{fig:transforms_summary}
\end{figure}

We simulate distribution shifts and corrupted data by applying transforms to images from a clean test set (\autoref{fig:methodology_overview}). We refer to the \textit{transformed} test set as a benchmarking dataset. As shown in \autoref{fig:transforms_summary}, we selected 11 different transforms that cover distribution shifts across multiple categories discussed previously: (1) signal-to-noise ratio (SNR); (2) contrast and intensity non-uniformity; (3) image resolution and blurring; (4) spatial location of features (e.g., translations, rotations, deformations); (5) presence of motion artifacts. Ground truth segmentations were transformed accordingly and re-discretized for spatial transforms that alter the shape or orientation of the feature of interest (e.g., affine, elastic deformation, downsampling). Transforms were sourced from the Medical Open Network for AI (MONAI)\footnote{\url{https://monai.io/}} and TorchIO\footnote{\url{https://torchio.readthedocs.io/}} \citep{perez2021torchio} libraries, both part of the PyTorch Ecosystem \citep{paszke2019pytorch}, or implemented where implementations were unavailable (e.g., for MRI [Rician] noise). See \ref{app:transforms} for full descriptions of each transform and their formulations.

\begin{figure}[t!]
\centering
\includegraphics[width=\textwidth]{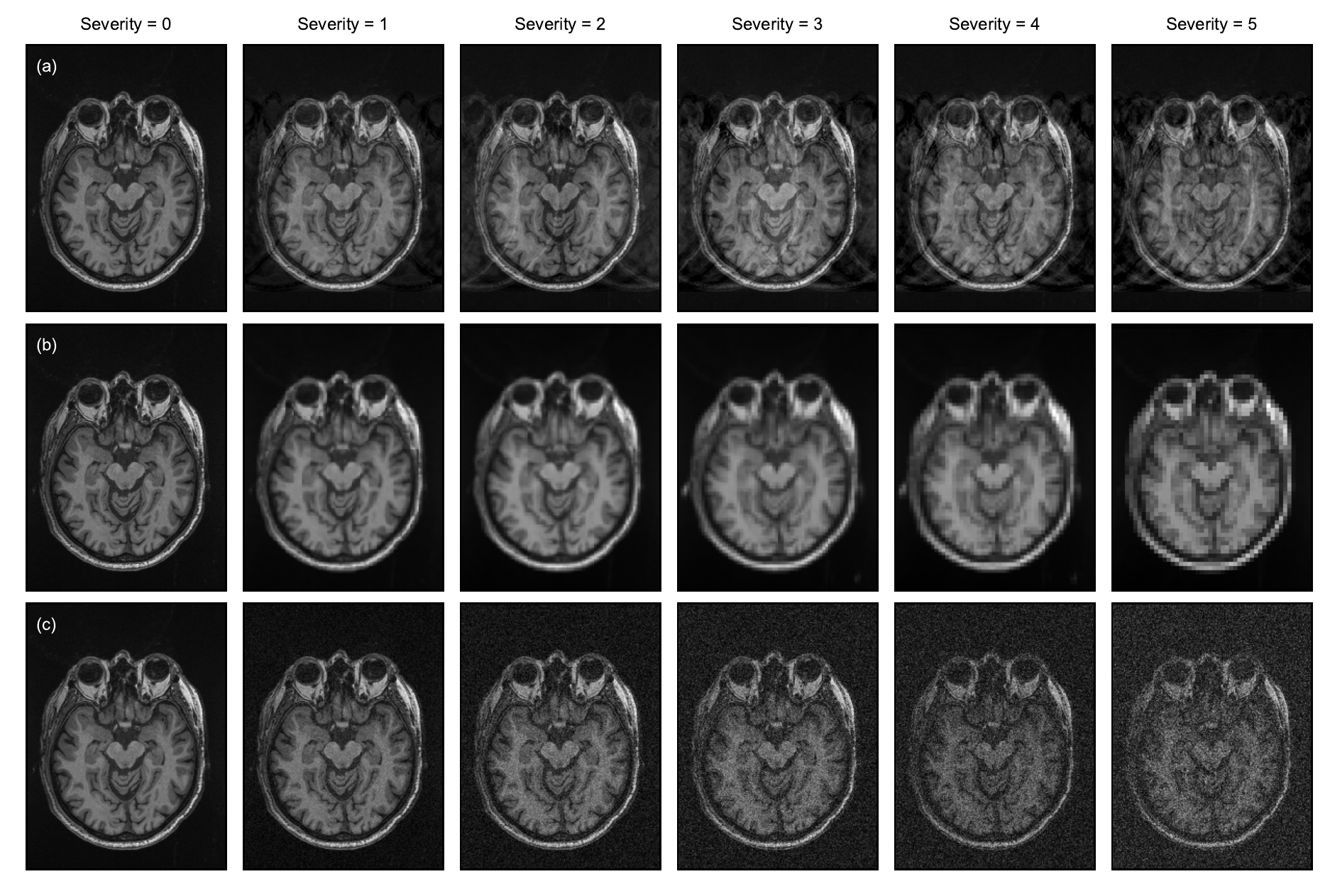}
\caption{Transform severity levels with increasing severity from left (severity $=$ 0; no transform applied) to right (severity $=$ 5; maximum level applied). Examples are shown for the (a) ghosting (motion artifact), (b) isotropic downsampling, and (c) MRI noise transforms. The images shown are taken from a T1 volume sliced axially through the hippocampus.}
\label{fig:severity_levels}
\end{figure}

Following the approach of \citet{hendrycks2019benchmarking}, we defined five distinct severity levels for each transform (see \autoref{fig:severity_levels}). Severity levels capture various magnitudes of distribution shifts, from mild to severe. Ideally, the chosen severity levels would span the range of distribution shifts seen in practice across imaging sites. However, there is a lack of quantitative studies investigating these ranges in the field, and it is not straightforward to translate observed variances into parametrically-modeled transforms. Thus, severity levels were chosen based on extensive visualization to fit in line with mild and extreme cases from the authors' experience with multi-site studies, and corruptions observed in practice (e.g., due to poor quality, artifacts or patient movement). To properly model severity levels and corruptions, scans with artifacts were removed or corrected in our clean test set prior to transformation. These severity levels may need to be re-optimized or calibrated if applied to other applications/tasks where the dataset characteristics differ drastically from those used in this study. For transform parameterizations corresponding to each severity level, see Supplementary Table S1.

\subsection{Tasks and datasets}
\label{sec:tasks_datasets}

To validate the utility of our methodology in benchmarking the robustness of DNNs to OOD data, we performed a series of experiments on three MR neuroimaging segmentation tasks, namely, segmentation of (1) both hippocampi, (2) the lateral ventricles, and (3) white matter hyperintensities (WMHs). Each of these tasks correspond to a critical structural imaging biomarker of neurodegeneration and cognitive decline, implicated in aging and a wide range of neurological disorders, including dementia, both clinically and in the research setting. Furthermore, we chose these three tasks for their diversity in difficulty and nature of the features of interest: while hippocampus and ventricle segmentation are relatively difficult and easy anatomical tasks, respectively, WMH segmentation is a challenging lesion-based task where the size, shape, and location of the features of interest vary significantly across subjects.

The distribution shifts modeled by transforms in our proposed benchmarking platform are generally applicable across the majority of MRI sequences. As such, we chose tasks using different MRI sequences as their primary modality: whole-brain T1-weighted MRI for hippocampus and ventricle segmentation, and whole-brain T2-weighted fluid-attenuated inversion recovery (FLAIR) MRI for WMH segmentation. FLAIR contrast significantly improves the detection of WMHs over T1-weighted scans \citep{bakshi2001fluid}.

\begin{table}[h!]
    \centering
    \caption{Dataset composition for the three tasks used in this study.}
    \vspace{2mm}
    \begin{threeparttable}
        \resizebox{0.5\textwidth}{!}{
        \begin{tabular}{llrrr}
        \hline
        Task (MRI sequence) & Study & Training & Test & Total \\
        \hline
        Hippocampus & ADNI & 107 & 28 & 135 \\
        segmentation & SDS & 79 & 20 & 99 \\
        (T1) & UPenn & 19 & 4 & 23 \\
         & \textbf{Total:} & 205 & 52 & 257 \\
        \hline
        Ventricle & CAIN & 281 & 0 & 281 \\
        segmentation & ONDRI & 214 & 82 & 296 \\
        (T1) & MITNEC & 0 & 50 & 50 \\
         & \textbf{Total:} & 495 & 132 & 627 \\
        \hline
        WMH segmentation & CAIN & 203 & 50 & 253 \\
        (FLAIR) & ONDRI & 160 & 38 & 198 \\
         & LIPA & 37 & 10 & 47 \\
         & VBH & 32 & 8 & 40 \\
         & MITNEC & 0 & 53 & 53 \\
         & \textbf{Total:} & 432 & 159 & 591 \\
        \hline
        \end{tabular}}
        \label{tab:datasets}
    \end{threeparttable}
\end{table}

The datasets for these tasks have been described in detail in previously published work \citep{goubran2020hippocampal, ntiri2021improved, mojirideep}. Each dataset comprises patient scans from multiple studies. This is representative of a typical model development scenario at clinical scale, where pooling from different sites allows for larger datasets to train highly parameterized deep learning models. Studies pooled together for the same task used roughly the same acquisition protocols for a given sequence. Acquisition details and parameters for each dataset are summarized in the supplementary materials.

\subsubsection{Hippocampus segmentation dataset}

The hippocampus segmentation dataset consists of 135 patients with Alzheimer's disease (AD), mild cognitive impairment (MCI) or healthy normal controls (NC) ($74.1 \pm 7.8$ years, 51\% male) from the Alzheimer's Disease Neuroimaging Initiative (ADNI) database \citep{boccardi2015training}, 99 individuals with AD, vascular cognitive impairment (VCI), or healthy normal controls ($85.3 \pm 11.2$ years, 56\% male) from the Sunnybrook Dementia Study (SDS) \citep{deshpande200419}, and 23 cases with temporal lobe epilepsy (TLE) from the University of Pennsylvania (UPenn) TLE atlas \citep{das2009structure}. Cases from each of these studies were used in both the training and test splits (\autoref{tab:datasets}).

\subsubsection{Ventricle segmentation dataset}

The ventricle segmentation dataset consists of 296 patients with cerebrovascular disease $\pm$ vascular cognitive impairment (CVD $\pm$ VCI) or Parkinson's disease (PD) (55-86 years, 75\% male)  from the multisite Ontario Neurodegenerative Disease Research Initiative (ONDRI)\footnote{\url{https://ondri.ca/}} \citep{farhan2017ontario, ramirez2020ontario}, and 281 individuals with non-surgical carotid stenosis (47-92, 61\% male) from the Canadian Atherosclerosis Imaging Network (CAIN) (\href{https://clinicaltrials.gov/}{ClinicalTrials.gov}: NCT01440296). 50 cases with severe WMH burden (Fazekas 3/3) were added to the test set from the Medical Imaging Trials Network of Canada (MITNEC) Project C6 as an unseen study (\href{https://clinicaltrials.gov/}{ClinicalTrials.gov}: NCT02330510).

\subsubsection{WMH segmentation dataset}

The WMH dataset consists of 253 patients with non-surgical carotid stenosis (47-92, 61\% male) from the CAIN, 198 patients with CVD $\pm$ VCI or PD (55-86, 75\% male) from ONDRI, 47 patients with nonfluent progressive aphasia, semantic dementia (SD), or normal healthy controls (55-80) from the Language Impairment in Progressive Aphasia (LIPA) study \citep{marcotte2017white}, and 40 patients with AD, CVD, or VCI (46-78, 50\% male) from the Vascular Brain Health (VBH) study \citep{swardfager2017peripheral}. 53 cases from the MITNEC Project were also added to the WMH test set as an unseen study.

\subsection{Metrics}
\label{sec:metrics}

\subsubsection{Segmentation metrics}
\label{sec:seg_metrics}

We use two metrics to highlight unique characteristics about the similarity between a segmentation prediction and the ground truth. To quantify the overlap, we use the \textit{Dice similarity coefficient (DSC)} \citep{taha2015metrics}. Given two sets $A$ and $B$ (e.g., a binary segmentation prediction and the ground truth), the DSC is defined as
\begin{equation}
    \mathrm{DSC} = \frac{2|A \cap B|}{|A| + |B|},
    \label{eq:dice}
\end{equation}
where $|A \cap B|$ is the cardinality of the intersection between $A$ and $B$, and $|A|$ and $|B|$ are the cardinalities of the two sets themselves. Working with Boolean data, the DSC can also be defined as $\mathrm{DSC} = 2\mathrm{TP} / (2\mathrm{TP} + \mathrm{FP} + \mathrm{TN})$, where $\mathrm{TP}$, $\mathrm{FP}$, and $\mathrm{TN}$ are the number of true positive, false positive, and true negative voxels, respectively, in a segmentation prediction relative to the ground truth. 

As a distinct but complementary metric, we use the \textit{Hausdorff distance} to quantify the greatest distance from a point marked as positive in the prediction to the closest point marked as positive in the ground truth, or vice versa. Traditionally, the Hausdorff distance between two sets of points $A$ and $B$ is defined as
\begin{equation}
    d_H = \max\bigg\{\max_{a \in A}\min_{b \in B}d(a,b), \max_{b \in B}\min_{a \in A}d(a,b)\bigg\},
    \label{eq:hausdorff}
\end{equation}
where $d(a, b)$ represents the Euclidean distance between two points $a \in A$ and $b \in B$. We modified the traditional definition by substituting the 95th percentile instead of the maximum in the max-min operation. The modified metric is often referred to as the \textit{modified (95th-percentile) Hausdorff distance (HD95)}. Both the $\mathrm{DSC}$ and $\mathrm{HD95}$ are used ubiquitously throughout the medical image segmentation literature \citep{taha2015metrics}.

\subsubsection{Benchmarking and robustness metrics}
\label{sec:benchmarking_metrics}

We derive several new metrics for benchmarking models based on robustness to distribution shifts in the context of medical image segmentation. While taking inspiration from \citet{hendrycks2019benchmarking} and \citet{kamann2020benchmarking} in computer vision, we build on their metrics in some important ways. First, we base our benchmarking metrics on \textit{multiple} segmentation metrics that highlight unique characteristics about the similarity between segmentation predictions and ground truth (see Section \ref{sec:seg_metrics}). Second, we quantify how the mean \textit{and} variance of segmentation metrics change in response to a transformed test set to capture additional insights regarding how DNN performance degrades with distribution shift. Third, we introduce weighting across severity levels to reflect the lower likelihood of seeing extreme cases of distribution shift in practice. Lastly, we eliminate the need for a reference model to calculate benchmarking metrics (simplifying the benchmarking procedure). Thus, a model can be tested and benchmarked in isolation, without needing to train or acquire another reference model.

We now introduce the first group of metrics which capture overall performance (as per the clean test set), but with a correction accounting for OOD data. Let $\mathrm{mDSC}_{T,s}$ and $\mathrm{sDSC}_{T,s}$ be the mean and standard deviation, respectively, of the DSC for a given model across an entire test set transformed by transform $T$ at severity level $s$. HD95 metrics are defined analogously (i.e., $\mathrm{mHD95}_{T,s}$ and $\mathrm{sHD95}_{T,s}$). We define the \textit{weighted mean DSC} for a given model and transform $T$ as
\begin{equation}
    \mathrm{wmDSC}_T = \frac{1}{\sum_{s=0}^5 w_s} \sum_{s=0}^5 w_s \cdot \mathrm{mDSC}_{T,s},
    \label{eq:wmDSC}
\end{equation}
where $s = 0$ represents the clean test set; we introduce $w_s = \alpha^s$, $0 < \alpha \leq 1$, as a weighting function across severity levels. Similarly, we define the \textit{weighted DSC standard deviation},
\begin{equation}
    \mathrm{wsDSC}_T = \frac{1}{\sum_{s=0}^5 w_s} \sum_{s=0}^5 w_s \cdot \mathrm{sDSC}_{T,s},
    \label{eq:wsDSC}
\end{equation}
using the same weighting function. HD95 metrics are defined analogously\footnote{Note that null (empty) segmentation predictions result in an infinite HD95. Thus, any null predictions are excluded from the calculations of $\mathrm{mHD95}_T$ and $\mathrm{sHD95}_T$. Across the results, we include the number of null predictions generated by a model for a given test set to complement the interpretation of these metrics.} (i.e., $\mathrm{wmHD95}_T$ and $\mathrm{wsHD95}_T$). These four weighted metrics constitute the first group of benchmarking metrics we present. They can be interpreted as offering a correction to their clean counterparts (e.g., $\mathrm{mDSC}_{\mathrm{clean}}$) by accounting for OOD data modeled by $T$. The degree of correction can be controlled by the weighting parameter, $\alpha$, with $\alpha = 1$ resulting in equal weighting across all severity levels (including the clean test set), and $\alpha \rightarrow 0$ approaching no correction to the clean metric. The choice of an exponential function implies that each severity level is weighted ``$\alpha$'' times as strongly as the previous severity level. We use $\alpha = 2/3$ as a baseline unless otherwise specified. Ideally, the weighting function would reflect the distribution of severity levels seen in practice across imaging sites (see limitations discussed in Section \ref{sec:dsc:limitations}).

We also introduce a group of metrics that are relatively agnostic to performance on the clean test set, to quantify the \textit{robustness} to a given transform $T$. We define robustness as how little model performance degrades \textit{relative} to performance on the clean test set. First, we define the \textit{mean-based DSC degradation},
\begin{equation}
    \mathrm{mDDeg}_T = \frac{1}{\sum_{s=1}^5 w_s} \sum_{s=1}^5 w_s (\mathrm{mDSC}_{\mathrm{clean}} - \mathrm{mDSC}_{T,s}),
    \label{eq:mDDeg}
\end{equation}
using the same weighting function as above. Note that $\mathrm{mDDeg}_T$ can be negative if $\mathrm{mDSC}$ improves on transformed data. We also define the \textit{variance-based DSC degradation},
\begin{equation}
    \mathrm{vDDeg}_T = \frac{1}{\sum_{s=1}^5 w_s} \sum_{s=1}^5 w_s (\mathrm{sDSC}_{T,s} - \mathrm{sDSC}_{\mathrm{clean}}),
    \label{eq:vDDeg}
\end{equation}
assuming that variance in $\mathrm{DSC}$ across a transformed test set is generally higher than across the clean test set. We define robustness metrics for $\mathrm{HD95}$ analogously:
\begin{equation}
    \mathrm{mHDeg}_T = \frac{1}{\sum_{s=1}^5 w_s} \sum_{s=1}^5 w_s (\mathrm{mHD95}_{T,s} - \mathrm{mHD95}_{\mathrm{clean}}),
    \label{eq:mHDeg}
\end{equation}
\begin{equation}
    \mathrm{vHDeg}_T = \frac{1}{\sum_{s=1}^5 w_s} \sum_{s=1}^5 w_s (\mathrm{sHD95}_{T,s} - \mathrm{sHD95}_{\mathrm{clean}});
    \label{eq:vHDeg}
\end{equation}
noting that a lower $\mathrm{HD95}$ is preferred. Across all benchmarking metrics, we additionally drop the subscript to produce an aggregate measure for all transforms, e.g., $\mathrm{wmDSC} = \sum_{T \in \mathcal{T}} \mathrm{wmDSC}_T$, where $\mathcal{T}$ is the space of all transforms being considered.

\subsection{Network architectures and training protocols}
\label{sec:architectures_and_training}

At the time of writing, most modern architectures for medical image segmentation are fully convolutional and based on the seminal U-Net \citep{ronneberger2015u}. Various features have been added to the core U-Net structure over the years, resulting in improvements across a wide variety of tasks. Examples of such features include residual connections \citep{he2016deep}, novel attention gates \citep{schlemper2019attention}, recurrent convolutional layers \citep{alom2019recurrent}, nested structure \citep{zhou2019unet++}, and deep supervision \citep{isensee2019automated}, to name a few. While each of these networks still use convolution-based processing as the backbone of both the encoder and decoder, recently, Vision Transformers (ViTs) \citep{dosovitskiy2020image} have been proposed as a replacement for the fully convolutional encoder paradigm that has been dominant recently \citep{hatamizadeh2021unetr}. In this work, we compare these two fundamentally different processing paradigms based on robustness to OOD data. We opted towards using architectures and processing techniques (including loss functions and pre-processing) that represent the current state-of-the-art while allowing for easy extension of experiments, reproducibility, and minimization of confounding effects.

\subsubsection{Network architectures}
\label{sec:net_architectures}

\paragraph{Baseline U-Net}

As a baseline model, we used a 3D residual U-Net \citet{kerfoot2018left}. The model features an encoder-decoder pathway with skip connections that concatenate same-resolution features from the encoder to the decoder \citep{ronneberger2015u}. The encoder uses 2-strided convolutions with $3 \times 3 \times 3$ volumetric kernels to downsample the image, starting with 16 initial feature maps and doubling at each encoder step until reaching 256 feature maps at the bottom of the network. The decoder path uses 2-strided transposed convolutions with the same kernel size to upsample feature maps supplied by the encoder back to the original image volume size. The residual block in each encoder step consists of two 3D convolution layers, each followed by batch normalization, activation, and dropout \citep{srivastava2014dropout}. Parametric ReLU \citep{he2015delving} was used as the activation function throughout the network and dropout was used at a rate of 10\% (empirically determined to give best results on clean data).

\paragraph{UNEt TRansformer}

To compare with the baseline U-Net, we trained UNEt TRansformer (UNETR) models proposed by \citet{hatamizadeh2021unetr}. In comparison to the fully convolutional U-Net, UNETR uses a sequence-to-sequence, self attention-based processing mechanism in the encoder: 3D mini-patches are extracted from the input image (or patch), flattened, and projected into a $K$-dimensional feature space by a learned linear layer; this sequence of mini-patch embeddings is then fed into a stack of transformer blocks \citep{dosovitskiy2020image} to produce a series of encoder features with the same shape as the input mini-patch embedding sequence. Taking inspiration from the U-Net, these flattened features are reshaped (volumetrically) and upsampled using deconvolution blocks, joining via concatenation with lower-level upsampled features from the decoder, which follows a similar structure to the U-Net. As in \citet{hatamizadeh2021unetr}, we used $16 \times 16 \times 16$ as a mini-patch size to build sequences for the transformer encoder\footnote{Note that the mini-patch size is different from the patch size used in pre-processing.}, $K = 768$, and 12 stacked transformer blocks in the encoder. We used $768$ as an intermediate multilayer perceptron dimension size, as this was empirically found to produce the best results on our clean datasets. Additionally, a dropout rate of 10\% was used throughout the network.

\subsubsection{Pre-processing}
\label{sec:pre-processing}

Prior to training and benchmark dataset generation, all image volumes (training and test splits) were bias field-corrected using the N4 algorithm \citep{tustison2010n4itk}. Additionally, all scans from each study dataset were manually reviewed by imaging analysts for motion and other artifacts; volumes with artifacts were excluded from further processing (training and testing). Our \textit{online} pre-processing steps for all networks (i.e., built into the data loaders for training and evaluation) included resampling to 1 mm isotropic voxel size, re-orientation to the RAS+ coordinate space, and intensity normalization using Z-score normalization on each individual image volume.

\subsubsection{Data augmentation}
\label{sec:data_augmentation}

Data augmentation was performed \textit{online} during training; that is, augmentation transforms were applied to the data each time an image volume was loaded during training. The only augmentation transform applied during the training of all models was random flipping in all three spatial axes with a probability $p = 0.5$.

Patch-based models included a random patch sampling transform applied to the data during training, which may be interpreted as a form of data augmentation. Whenever patch sampling was applied to an image volume, four volumetric patches of size $P \times P \times P$ were randomly sampled from the image and collated along the batch dimension of the input tensor. Unless stated otherwise, $P = 96$ was used as the default patch size for all models. Sampled patches were centered on a foreground (positively labeled) voxel 50\% of the time, and a nonzero background voxel (e.g., brain tissue not corresponding to the feature of interest) 50\% of the time. At test time, patch-based models performed inference on whole image volumes by making segmentation predictions on sliding patches of the same size as those used during training with 25\% overlap between patches.

\subsubsection{Training}
\label{sec:training}

For each task, approximately 15\% of the training set was held out as a validation set. None of the samples in the validation sets were used at any point during training for calculating the loss function and backpropagating gradients through the network. Throughout training, the model was evaluated every two training epochs on the entire validation set, and the mean $\mathrm{DSC}$ was recorded. Each time a new best mean validation $\mathrm{DSC}$ was recorded, model parameters were saved, overwriting the previously saved model. In this sense, the final model obtained from training had the highest mean validation $\mathrm{DSC}$ over the total number of training epochs. Additionally, early stopping was set at 125 epochs for each model such that if the mean validation $\mathrm{DSC}$ did not improve in 125 epochs, training was stopped. U-Net and UNETR models were trained for a maximum of 500 and 1000 epochs, respectively, as these were deemed sufficient for each model to converge based on visualization of training curves.

Each model was trained using a Dice-based loss \citep{milletari2016v}, averaged over foreground- and background-labeled voxels. This loss formulation has been shown to alleviate the problem of class imbalance in medical image segmentation tasks. Inclusion of foreground and background loss terms was empirically found to yield best results on clean data in our previous work. Models were trained using the ADAM optimizer \citep{kingma2014adam} with a learning rate of $1 \times 10^{-4}$, which was empirically found to give best results on clean data. Batch size was controlled based on the type of model: whole-image or patch-based. Due to GPU memory constraints, whole-image models were trained using a batch size of one (in this case, batch normalization reduces to instance normalization), and patch-based models were trained using a batch size of 8 (including 4 sampled patches each from two volumes).

\subsubsection{Implementation}
\label{sec:implementation}

All networks were implemented using the Medical Open Network for AI (MONAI) library\footnotetext{\href{https://monai.io/}{https://monai.io/}} and trained on V100-SXM2 graphics cards with 32GB of memory and a Volta architecture (NVIDIA, Santa Clara, CA). We make our code for benchmark dataset generation, evaluation, and metric implementations, as well as the released hippocampus benchmarking dataset, publicly available at \url{https://github.com/AICONSlab/roodmri}.

\section{Experiments and results}

\subsection{Sensitivity to transforms: baseline U-Net}
\label{sec:baseline_sensitivity}

\begin{figure}[t!]
\centering
\includegraphics[width=\textwidth]{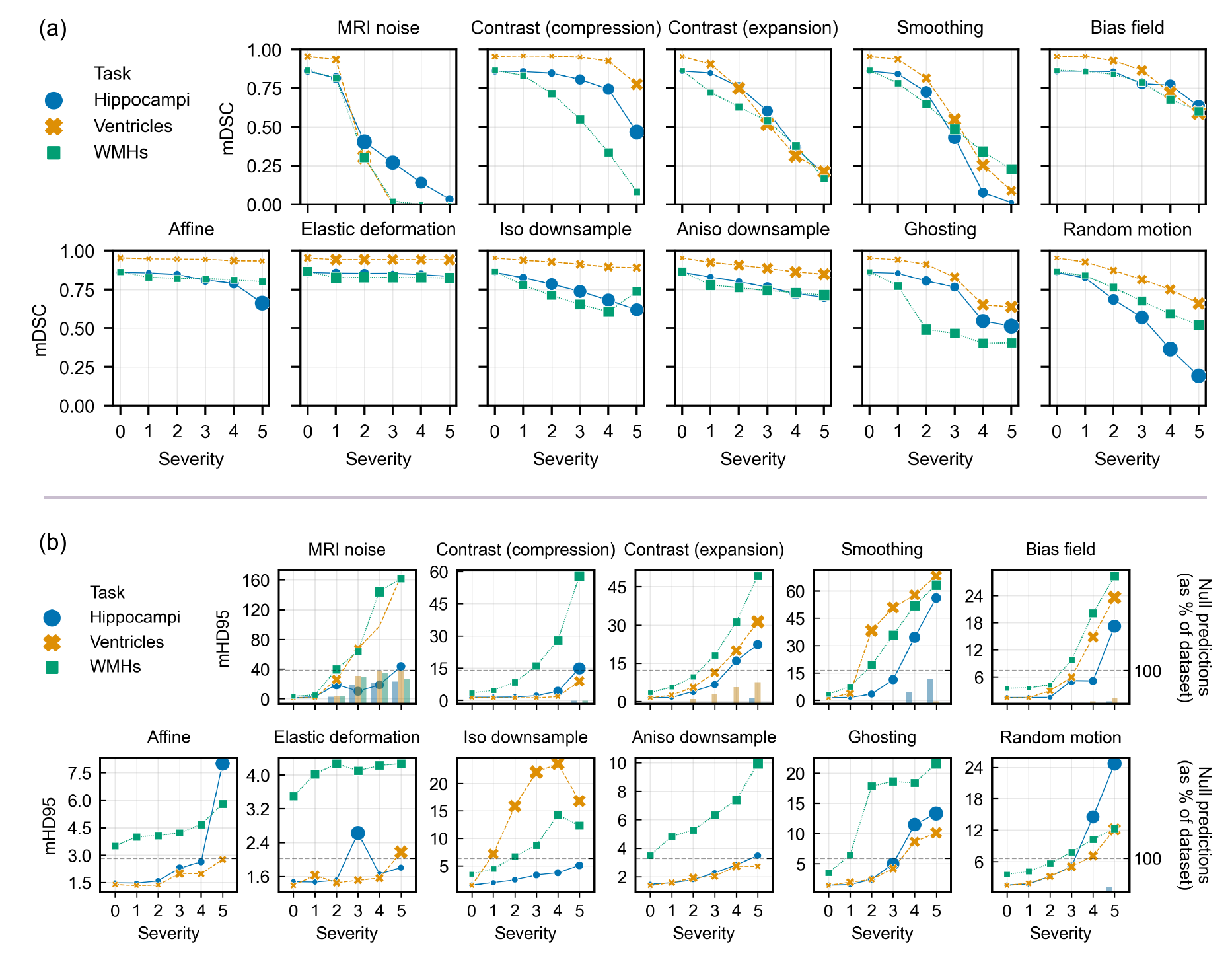}
\caption{Sensitivity curves for the baseline residual U-Net model to MR image transforms at varying degrees of severity, measured by (a) the mean Dice coefficient ($\mathrm{mDSC}$) and (b) modified (95th-percentile) Hausdorff distance ($\mathrm{mHD95}$) across each transformed test set. The sensitivity curves are presented for the U-Net model across all 11 transforms for each of the three tasks (hippocampi: blue circle; ventricles: orange cross; WMHs: green square). In each plot, marker size (area) is proportional to the metric standard deviation across the transformed test set. Since null (empty) predictions are not taken into account when computing the mHD95 metric, the percentage of null predictions on each dataset are given by the bars in (b) relative to the right axes.}
\label{fig:baseline_sensitivity_curves}
\end{figure}

\begin{figure}[t!]
\centering
\includegraphics[width=\textwidth]{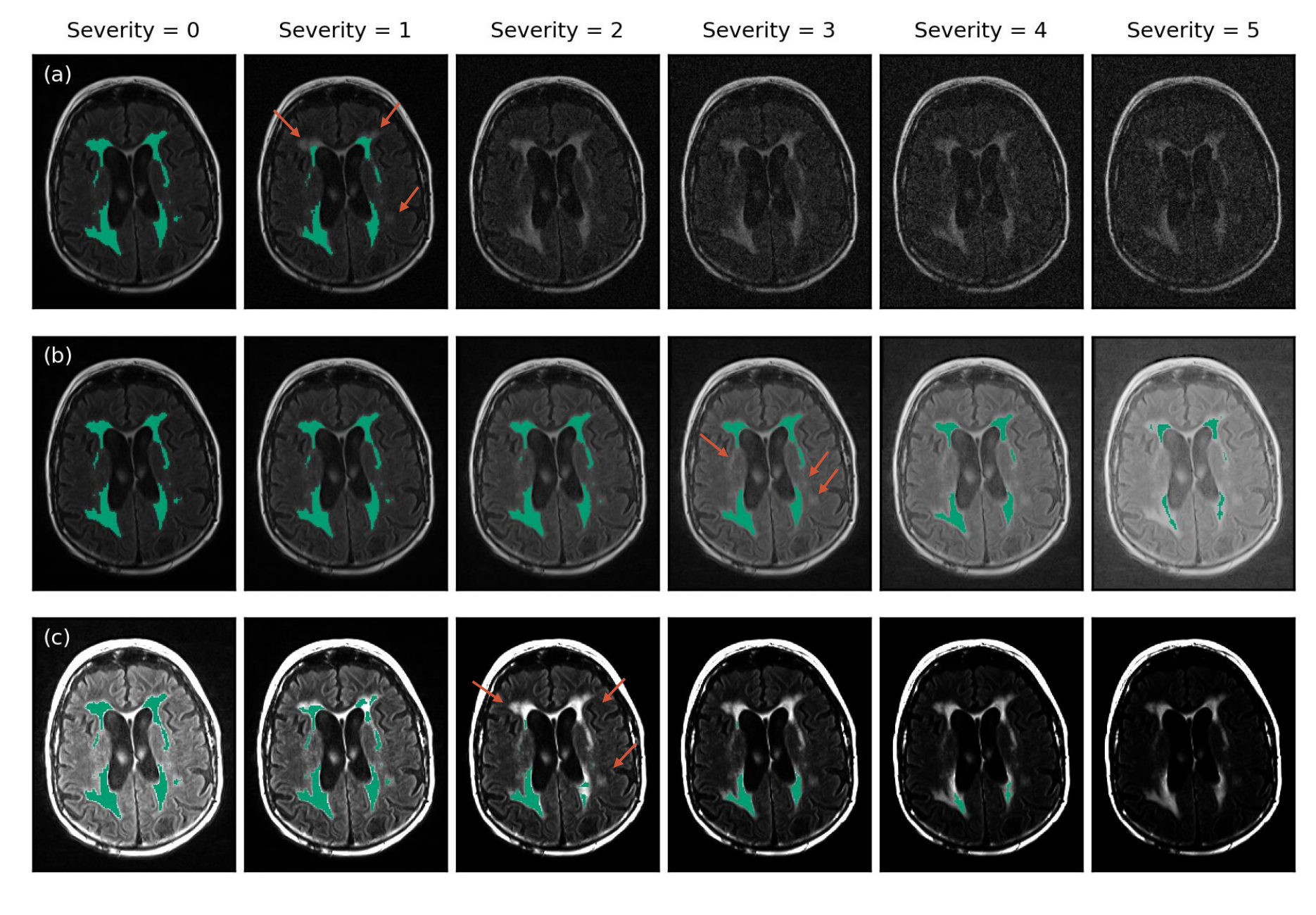}
\caption{WMH segmentation degradation over severity levels for three intensity-based transforms: (a) MRI noise; (b) gamma compression; (c) gamma expansion applied to an example subject. The same residual U-Net model was used to generate each prediction on the patient sample shown here (same FLAIR slice in each subfigure). Arrows indicate notable missed lesions from one severity level to the next.}
\label{fig:prediction_degradation_examples}
\end{figure}

We first assessed the sensitivity of the baseline U-Net architecture (described in Section \ref{sec:net_architectures}) to all transforms (summarized in Section \ref{sec:transforms}). Degradation metrics for each transform and task are provided in \autoref{tab:baseline_transform_metrics}, calculated using equations \ref{eq:mDDeg}-\ref{eq:vHDeg}. In Figure \ref{fig:baseline_sensitivity_curves}, sensitivity curves demonstrate that performance susceptibility varies greatly across transforms. For example, MRI (Rician) noise presents an especially strong effect influencing the residual U-Net's performance degradation. As the noise level ($\sigma_g/\sigma_{\mathrm{img}}$) increases from 0.16 to 0.32, $\mathrm{mDSC}$ drops by at least 50\% and $\mathrm{mHD95}$ increases 10-fold across all three tasks. The abrupt degradation between severity levels 1 and 2 is demonstrated qualitatively in \autoref{fig:prediction_degradation_examples}a for a subject with large WMH burden from the MITNEC dataset. While a trained human reader can segment WMH lesions all the way up to severity level 5 ($\sigma_g/\sigma_{\mathrm{img}} = 0.80$), the residual U-Net model fails to make any positive predictions on the slice shown in the figure after severity level 2 ($\sigma_g/\sigma_{\mathrm{img}} \geq 0.32$). \autoref{fig:prediction_degradation_examples}b-c demonstrate a more gradual degradation in segmentation predictions due to changes in contrast on the same subject. When the image is transformed using gamma compression, the model begins to miss small lesions by severity levels 2 and 3 ($\gamma = [0.72, 0.58]$); by severity level 5 ($\gamma = 0.30$), only the central portions of large periventricular WMH lesions remains properly segmented, with the remaining bordering regions being misclassified as normal-appearing white matter. Gamma compression is one of the only transforms where a marked difference in performance degradation was noticed across tasks. For WMH segmentation, $\mathrm{mDDeg}_{\mathrm{comp.}} = 0.22$ (compared to 0.06 and 0.02 for the hippocampus and ventricles, respectively) and $\mathrm{mHDeg}_{\mathrm{comp.}} = 10.7$ (compared to 1.5 and 0.5 for the hippocampus and ventricles, respectively). This difference is most likely due to the fact that WMHs appear as hyperintense on FLAIR and that gamma compression serves to decrease the relative contrast between hyperintense features while expanding the contrast between hypointense features in an image (ventricles, for instance, appear as hypointense to surrounding parenchyma on T1). WMH segmentation predictions degrade due to gamma expansion at a rate similar to the other tasks. \autoref{fig:prediction_degradation_examples}c illustrates that despite WMH lesions being more pronounced due to gamma expansion, the model still fails to classify the WMH voxels at higher levels of transform severity. In addition to a degradation in the central tendency (mean), the variance in $\mathrm{DSC}$ and $\mathrm{HD95}$ across the test set also increases with increasing severity of the transform applied (for example, gamma compression and random motion in \autoref{fig:baseline_sensitivity_curves}). In general, mean-based and variance-based robustness metrics exhibited a correlation at the transform level (Supplementary Figure S1).

\begin{table}[h!]
    \centering
    \caption{Degradation metrics for the baseline residual U-Net across all transforms, calculated with weighting across severity levels using $\alpha = 2/3$. The average metric value across transforms is provided in the rightmost column.}
    \vspace{2mm}
    \begin{threeparttable}
        \resizebox{\textwidth}{!}{
        \begin{tabular}{llllllllllllll}
        \hline
        Metric &       Task &  Noise &  Comp. &  Exp. &  Smooth. &   BF &  Affine &   ED &   ID &   AD &  Ghost. &   RM &  Avg. \\
        \hline
         mDDeg & Hippocampi &   0.38 &   0.06 &  0.18 &     0.27 & 0.04 &    0.04 & 0.01 & 0.09 & 0.07 &    0.09 & 0.21 &  0.13 \\
               & Ventricles &   0.52 &   0.02 &  0.27 &     0.26 & 0.08 &    0.01 & 0.01 & 0.03 & 0.05 &    0.09 & 0.10 &  0.13 \\
               &        WMHs &   0.47 &   0.22 &  0.28 &     0.26 & 0.06 &    0.04 & 0.04 & 0.15 & 0.11 &    0.29 & 0.13 &  0.19 \\
        \hline
         vDDeg & Hippocampi &   0.19 &   0.06 &  0.09 &     0.06 & 0.06 &    0.02 & -0.01 & 0.07 & 0.01 &    0.07 & 0.07 &  0.06 \\
               & Ventricles &   0.08 &   0.00 &  0.25 &     0.09 & 0.11 &   -0.02 &  0.02 & 0.04 & 0.03 &    0.07 & 0.05 &  0.07 \\
               &        WMHs &   0.06 &   0.07 &  0.10 &     0.09 & 0.04 &    0.01 &  0.01 & 0.07 & 0.02 &    0.13 & 0.05 &  0.06 \\
        \hline
         mHDeg & Hippocampi &   11.8 &    1.5 &   4.8 &     10.2 & 2.3 &     0.8 & 0.3 &  1.3 & 0.6 &     2.9 & 4.4 &   3.7 \\
               & Ventricles &   41.1 &    0.5 &   7.5 &     30.3 & 4.4 &     0.2 & 0.2 & 13.1 & 0.6 &     2.4 & 2.7 &   9.4 \\
               &        WMHs &   48.3 &   10.7 &  11.5 &     21.1 & 5.1 &     0.8 & 0.6 &  4.0 & 2.4 &    10.5 & 2.9 &  10.7 \\
        \hline
         vHDeg & Hippocampi &   13.3 &    3.2 &   6.3 &      6.4 & 5.6 &     1.3 & 1.3 &  2.3 & 0.2 &     5.3 & 4.2 &   4.5 \\
               & Ventricles &    8.1 &    0.6 &  14.1 &     18.5 & 9.4 &     0.5 & 2.2 & 24.4 & 2.0 &     3.9 & 4.6 &   8.0 \\
               &        WMHs &   18.4 &    7.1 &   7.4 &     14.1 & 5.7 &     0.9 & 0.9 &  7.7 & 2.4 &     3.7 & 1.9 &   6.4 \\
        \hline
        \end{tabular}}
        \begin{tablenotes}[flushleft]
            \footnotesize
            \item mDDeg: mean-based DSC degradation; vDDeg: variance-based DSC degradation; mHDeg: mean-based Hausdorff
            \item degradation; vHDeg: variance-based Hausdorff degradation; Comp.: contrast (gamma) compression; Exp.: contrast (gamma)
            \item expansion; Smooth.: smoothing; BF: bias field; ED: elastic deformation; ID: isotropic downsampling; AD: anisotropic
            \item downsampling; Ghost.: ghosting artifacts; RM: random motion artifacts.
        \end{tablenotes}
        \label{tab:baseline_transform_metrics}
    \end{threeparttable}
\end{table}

\begin{figure}[t!]
\centering
\includegraphics[width=\textwidth]{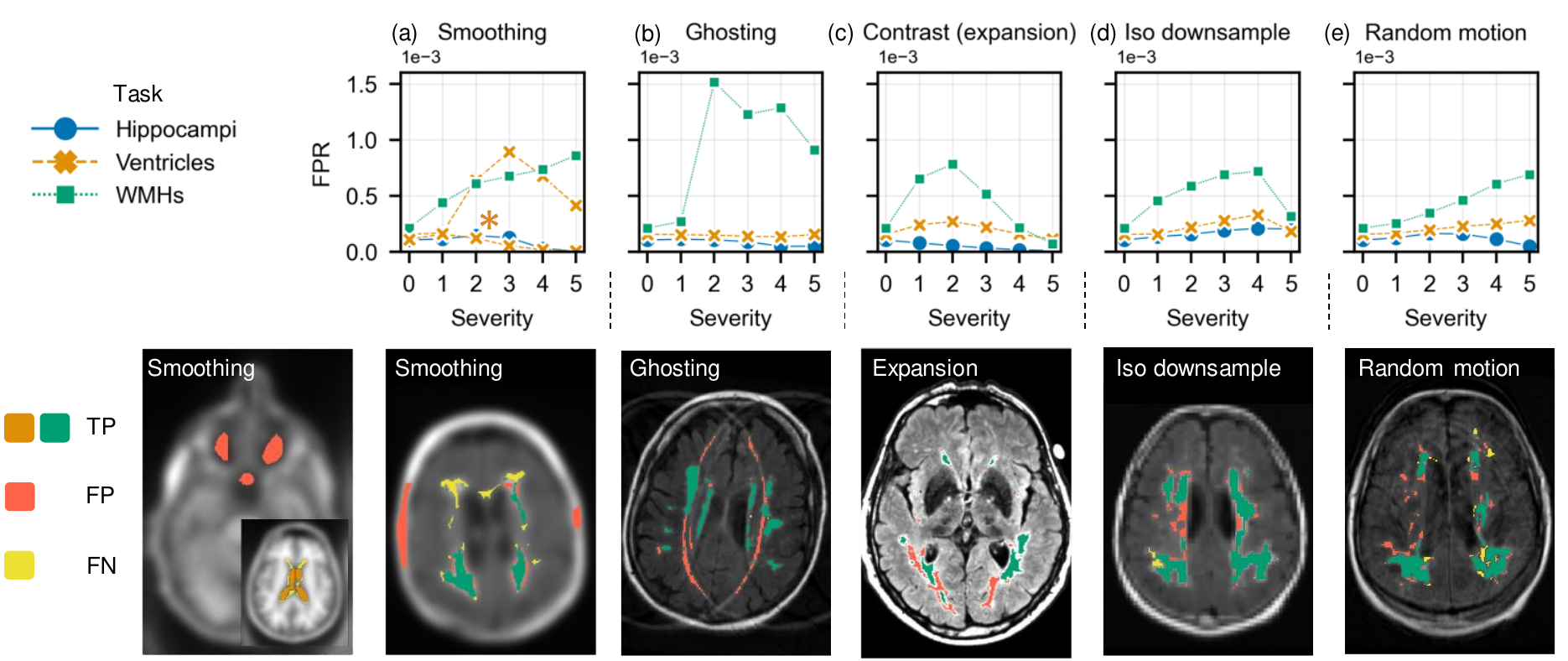}
\caption{False positive rate ($\mathrm{FPR}$) (top row) and corresponding segmentation predictions (bottom row) induced by transforms: (a) smoothing; (b) ghosting artifacts; (c) gamma expansion; (d) isotropic downsampling; (e) random motion artifacts. False positive rate ($\mathrm{FPR}$) curves are presented for each of the three tasks (hippocampi: blue circle; ventricles: orange cross; WMHs: green square). True positives ($\mathrm{TP}$) (orange for ventricles and green for WMHs), false positives ($\mathrm{FP}$) (red) and false negatives ($\mathrm{FN}$) (yellow) are referenced in the lower-left legend. All curves correspond to the patch-based residual U-Net model (patch size $96 \times 96 \times 96$ mm\textsuperscript{3}) described in Section \ref{sec:architectures_and_training}, with the exception of the asterisk-marked curve in subfigure (a) which was trained on whole image volumes.}
\label{fig:false_positives}
\end{figure}

While most of the transforms cause the model to miss positive segmentation predictions as shown in \autoref{fig:prediction_degradation_examples} (resulting in false negatives), some transforms have systematic ways of causing the model to generate false positive predictions (\autoref{fig:false_positives}). In \autoref{fig:false_positives}a, smoothing causes the U-Net to mistake the actual patient skull for WMH lesions while mistaking the maxillary and sphenoid sinuses for the ventricles. Interestingly, training the same network on whole image volumes as opposed to $96 \times 96 \times 96$ mm\textsuperscript{3} patches decreases the likelihood of this problem occurring for ventricle segmentation (marked by an asterisk in \autoref{fig:false_positives}a) but not WMH segmentation. Ghosting artifacts are another quintessential example: as shown in \autoref{fig:false_positives}b, the replicated skulls from ghosting artifacts (ghosts), if positioned accordingly, may be misclassified as periventricular WMH lesions. These ghost skulls that appear as hyperintense overlaid on the brain parenchyma can resemble the characteristic shape and intensity of WMHs which streak inward toward the occipital and frontal lobes. This effect was observed on numerous samples transformed with ghosting artifacts, being most prevalent when only two ghosts were present and their skulls aligned well with periventricular WMHs already present in the image. The other transforms observed to consistently induce false positive predictions were gamma expansion, isotropic downsampling, and random motion on the WMH segmentation task, examples of which are also shown in \autoref{fig:false_positives}.

\subsection{Effect of patch size on robustness to transforms across tasks}
\label{sec:patch_results}

\begin{table}[h!]
    \centering
    \caption{Benchmarking results of U-Nets trained with different patch sizes on both clean and transformed (corrupted) test sets for the three tasks. Weighted robustness metrics (e.g., $\mathrm{wmDSC}$) are aggregated across all 11 transforms applied to the test set. Best results across models are highlighted in bold font.}
    \vspace{2mm}
    \begin{threeparttable}
        \resizebox{\textwidth}{!}{
        \begin{tabular}{llllllllll}
        \hline
              Task &  Patch size (mm\textsuperscript{3}) &  $\mathrm{mDSC}_{\mathrm{clean}}$ &  $\mathrm{wmDSC}$ &  $\mathrm{sDSC}_{\mathrm{clean}}$ &  $\mathrm{wsDSC}$ &  $\mathrm{mHD95}_{\mathrm{clean}}$ &  $\mathrm{wmHD95}$ &  $\mathrm{sHD95}_{\mathrm{clean}}$ &  $\mathrm{wsHD95}$ \\
        \hline
        Hippocampi & Whole-image &        0.83 &   0.75 &        0.04 &   0.09 &          1.8 &     3.3 &          0.7 &     2.9 \\
         &          96 &        \textbf{0.86} &   0.78 &        0.05 &   0.09 &          1.5 &     3.8 &          0.5 &     3.4 \\
         &          64 &        \textbf{0.86} &   \textbf{0.79} &        \textbf{0.02} &   \textbf{0.06} &          \textbf{1.4} &     \textbf{3.0} &          \textbf{0.3} &     \textbf{2.3} \\
         &          32 &        0.85 &   0.76 &        0.03 &   0.07 &         10.4 &    14.8 &         22.6 &    23.9 \\
         \hline
        Ventricles & Whole-image &        \textbf{0.96} &   \textbf{0.87} &        0.08 &   0.11 &          \textbf{1.2} &     \textbf{6.5} &          1.6 &     \textbf{6.2} \\
         &          96 &        0.95 &   \textbf{0.87} &        0.05 &   0.09 &          1.4 &     7.3 &          \textbf{1.1} &     \textbf{6.2} \\
         &          64 &        0.95 &   0.86 &        \textbf{0.03} &   \textbf{0.08} &          2.3 &     9.2 &          6.2 &    11.5 \\
         &          32 &        0.90 &   0.77 &        0.07 &   0.11 &         31.8 &    43.8 &         29.8 &    29.2 \\
        \hline
               WMHs & Whole-image &        0.83 &   0.72 &        0.10 &   0.13 &          5.1 &    12.6 &          5.8 &    10.8 \\
                &          96 &        \textbf{0.86} &   \textbf{0.75} &        \textbf{0.08} &   \textbf{0.12} &          \textbf{3.5} &    \textbf{10.3} &          \textbf{5.1} &     \textbf{9.2} \\
                &          64 &        0.85 &   0.73 &        0.09 &   0.13 &          3.8 &    10.4 &          5.3 &     9.7 \\
                &          32 &        0.76 &   0.63 &        0.14 &   0.17 &         22.7 &    32.0 &         20.8 &    21.2 \\
        \hline
        \end{tabular}}
        \label{tab:patch_high-level_metrics}
    \end{threeparttable}
\end{table}

\begin{figure}[t!]
\centering
\includegraphics[width=\textwidth]{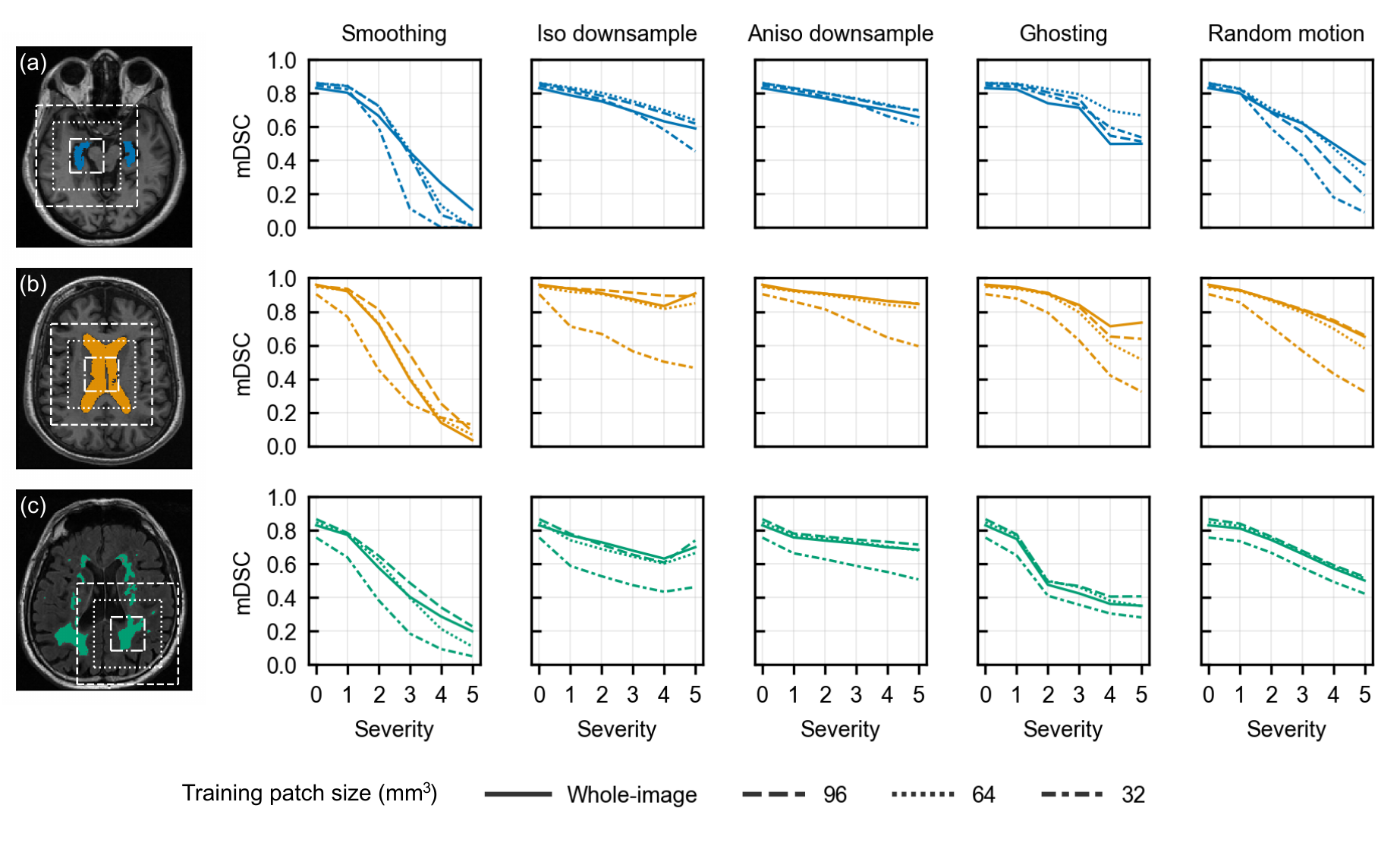}
\caption{Patch size visualizations (left panel) and degradation in $\mathrm{mDSC}$ across \textit{blurring} transforms for (a) hippocampus, (b) ventricle, and (c) WMH segmentation. Line types (straight, dashed, or dotted pattern) correspond to whole-image or different patch size models in both axial slice visualization and mDSC degradation curves.}
\label{fig:patch_size_sensitivity_curves}
\end{figure}

\begin{table}[h!]
    \centering
    \caption{Robustness of U-Nets trained with different patch sizes across \textit{blurring} transforms: smoothing, isotropic/anisotropic downsampling, ghosting, and random motion artifacts. Highest mean Dice degradation (i.e., worst results) across models are highlighted in bold font.}
    \vspace{2mm}
    \begin{threeparttable}
        \begin{tabular}{lllllll}
            \hline
             &  & \multicolumn{5}{l}{\underline{$\mathrm{mDDeg}_T$}} \\
            Task & Patch size (mm\textsuperscript{3}) & Smooth. & ID & AD & Ghost. & RM \\
            \hline
            Hippocampi & Whole-image & 0.24 & 0.10 & 0.07 & \textbf{0.11} & 0.16 \\
             & 96 & 0.27 & 0.09 & 0.07 & 0.09 & 0.21 \\
             & 64 & 0.26 & 0.08 & 0.07 & 0.06 & 0.18 \\
             & 32 & \textbf{0.36} & \textbf{0.12} & \textbf{0.08} & 0.09 & \textbf{0.29} \\
            \hline
            Ventricles & Whole-image & 0.34 & 0.06 & 0.06 & 0.09 & 0.11 \\
             & 96 & 0.26 & 0.03 & 0.05 & 0.09 & 0.10 \\
             & 64 & 0.31 & 0.06 & 0.06 & 0.11 & 0.11 \\
             & 32 & \textbf{0.42} & \textbf{0.27} & \textbf{0.12} & \textbf{0.18} & \textbf{0.22} \\
            \hline
            WMHs & Whole-image & 0.27 & 0.11 & 0.10 & 0.28 & 0.11 \\
             & 96 & 0.26 & 0.15 & 0.11 & \textbf{0.29} & \textbf{0.13} \\
             & 64 & 0.29 & 0.16 & 0.10 & 0.28 & 0.12 \\
             & 32 & \textbf{0.37} & \textbf{0.23} & \textbf{0.14} & \textbf{0.29} & 0.12 \\
            \hline
        \end{tabular}
        \label{tab:patch_DSC_degradations}
    \end{threeparttable}
\end{table}

\begin{figure}[t!]
\centering
\includegraphics{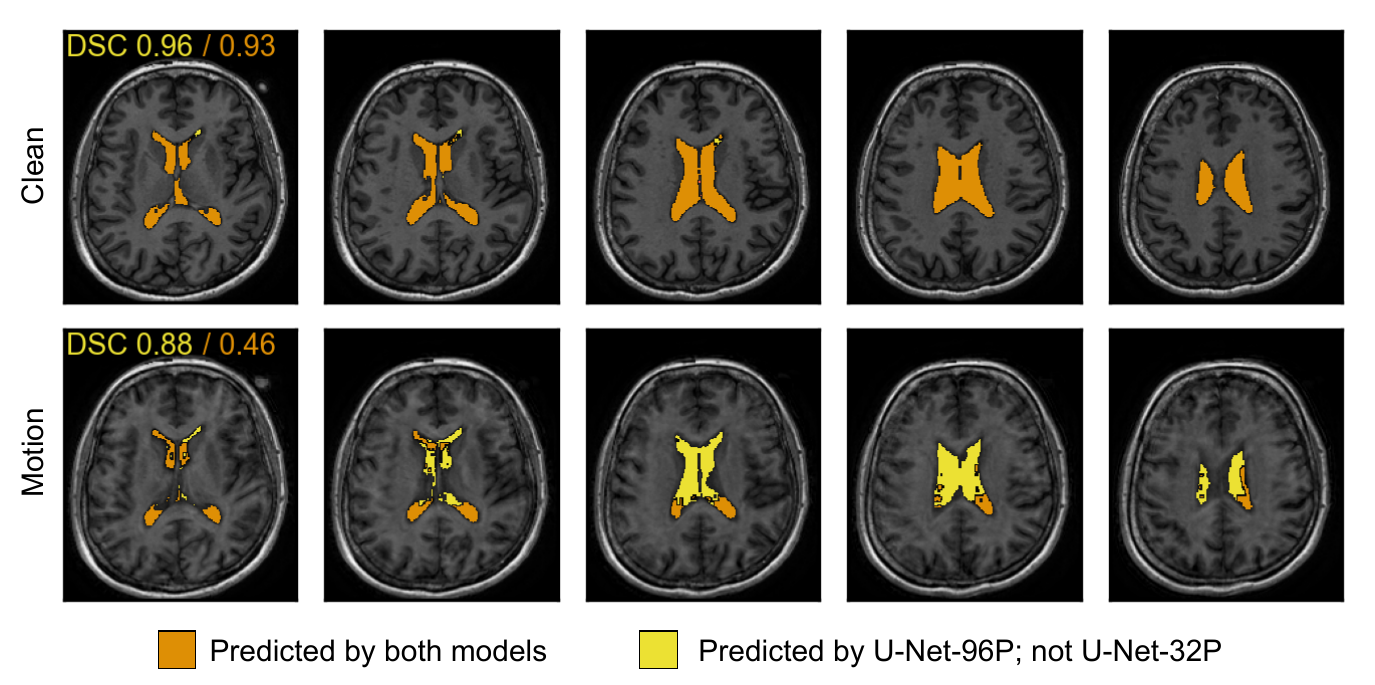}
\caption{Mild motion artifacts cause U-Nets trained with small patch sizes to miss segmentation predictions on large structures such as the lateral ventricles. Axial slices are shown 4 mm apart on the clean (no motion; top row) and motion-transformed (bottom row) T1 images of the same subject. $\mathrm{DSC}$ values are shown in the top-left corner of each row for a U-Net trained on $96 \times 96 \times 96$ mm\textsuperscript{3} patches (U-Net-96P; yellow) and a U-Net trained on $32 \times 32 \times 32$ mm\textsuperscript{3} patches (U-Net-32P; orange).}
\label{fig:small_patch_size_motion_degradation}
\end{figure}

Besides highlighting the sensitivity of modern DNNs to transformed/corrupted data in neuroimaging, one of the goals of this work was to provide a framework and metrics for benchmarking candidate models on the basis of robustness. To pilot our methodology, we compared four models trained on different patch sizes (shown in \autoref{fig:patch_size_sensitivity_curves}) across transformed datasets for the three tasks. The model architecture was identical to the baseline U-Net in Section \ref{sec:baseline_sensitivity}; the only difference between models being the patch size used for training and evaluation (see Section \ref{sec:architectures_and_training} for details). This training scheme limits the receptive field of the network in smaller-patch-size models. Benchmarking metrics aggregated across all transforms are tabulated in \autoref{tab:patch_high-level_metrics} for all three tasks. In many cases, patch-based models with a large enough patch size (e.g., $96 \times 96 \times 96$ mm\textsuperscript{3}) outperformed whole-image models, both on the clean test set, and on the basis of robustness across transformed test sets. Despite achieving comparable results to the other models in terms of $\mathrm{DSC}$ on the hippocampus segmentation task, models trained on $32 \times 32 \times 32$ mm\textsuperscript{3} patches (U-Net-32P) struggled comparatively with overlap-based metrics on ventricle and WMH segmentation, and displayed higher distance-based metric values on clean test sets across all transforms. Even relative to clean test set performance, the degradation across severity levels was far more pronounced for the U-Net-32P models for a subset of transforms which we refer to as \textit{blurring} transforms: smoothing, isotropic/anisotropic downsampling, and motion artifacts (ghosting and random motion) (\autoref{fig:patch_size_sensitivity_curves}; see Supplementary Figure S2 for HD95 curves). Moreover, this trend is present to some degree in all three tasks, although tasks with large feature-of-interest sizes, such as ventricle segmentation, appear to be more strongly affected (\autoref{fig:patch_size_sensitivity_curves} and \autoref{tab:patch_DSC_degradations}). \autoref{fig:small_patch_size_motion_degradation} illustrates this point, demonstrating ventricle segmentation predictions from U-Nets trained on $96 \times 96 \times 96$ mm\textsuperscript{3} (U-Net-96P) and $32 \times 32 \times 32$ mm\textsuperscript{3} (U-Net-32P) patches for a subject from the ONDRI cohort with and without synthetic motion artifact corruption. On the original (clean) image volume both models perform well, achieving a $\mathrm{DSC}$ of 0.96 and 0.93 for the U-Net-96P and U-Net-32P models, respectively. After the image has been corrupted by mild random motion (severity level 2) that is hardly noticeable without comparison to the clean data (representing corruption levels which are likely encountered in the clinical and research setting), the U-Net-96P model maintains a reasonable $\mathrm{DSC}$ of 0.88, while the 32P model's $\mathrm{DSC}$ drops substantially by almost 50\% to 0.46.

\subsection{Effect of data augmentation on robustness}
\label{sec:augmentation_results}

\subsubsection{Augmentation with a single transform}
\label{sec:augmentation_single}

\begin{figure}[t!]
\centering
\includegraphics[width=\textwidth]{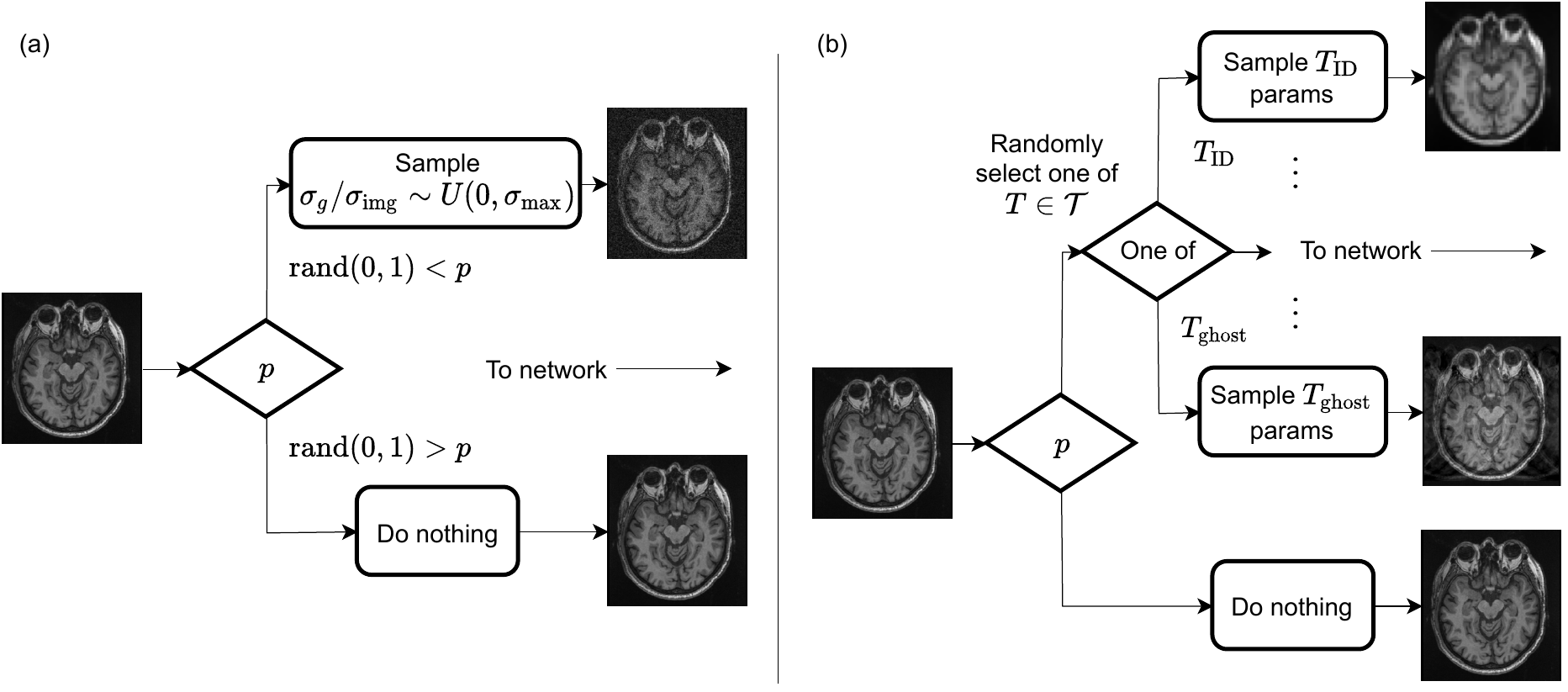}
\caption{Data augmentation schemes using (a) single and (b) multiple augmentation transforms. Transforms were applied with probability $p$ for each sample loaded during training. Transform parameters were sampled uniformly between the parameter value that constitutes no transformation (e.g., zero in the case of noise) and the parameter value constituting the maximum intensity of transformation desired. In the case of multiple transforms, one transform is randomly selected (with equal probability) each time augmentation is applied.}
\label{fig:augmentation_scheme}
\end{figure}

In this experiment, we investigated whether simple augmentation strategies can eliminate most of the sensitivities observed in Section \ref{sec:baseline_sensitivity}, without modifications to network architecture. We started by augmenting training samples using a single transform, as shown in \autoref{fig:augmentation_scheme}a (see Section \ref{sec:architectures_and_training} for training details). First, we studied if simply providing mildly (i.e., low-severity) transformed inputs to the network was enough to develop robustness to a particular transform across severity levels, or whether the intensity range of the augmentation transform applied during training had an effect on robustness at test time. As another key complementary variable in the augmentation scheme, we studied the frequency $p$ with which transforms were applied to the training data, and how this parameter affects robustness for varying degrees of transform intensity seen by the model through data augmentation. Second, we studied whether augmentation with a single transform can improve robustness to other transforms. We started by training the baseline U-Net studied in Section \ref{sec:baseline_sensitivity} using MRI noise as an augmentation transform, with varying probabilities of transformation during training, $p$, and with varying intensity ranges of noise supplied to the network during training. Intensity was controlled by sampling $\sigma_g/\sigma_{\mathrm{img}} \sim U(0, \sigma_{\mathrm{max}})$ uniquely for each newly loaded training sample, where $\sigma_{\mathrm{max}}$ controls the maximum intensity of noise applied to the image relative to the image intensity standard deviation. Models were trained with $\sigma_{\mathrm{max}} \in \{0.16, 0.32, 0.48, 0.64, 0.80\}$, corresponding to the noise severity levels in the benchmarking test sets.

\begin{figure}[t!]
\centering
\includegraphics[width=\textwidth]{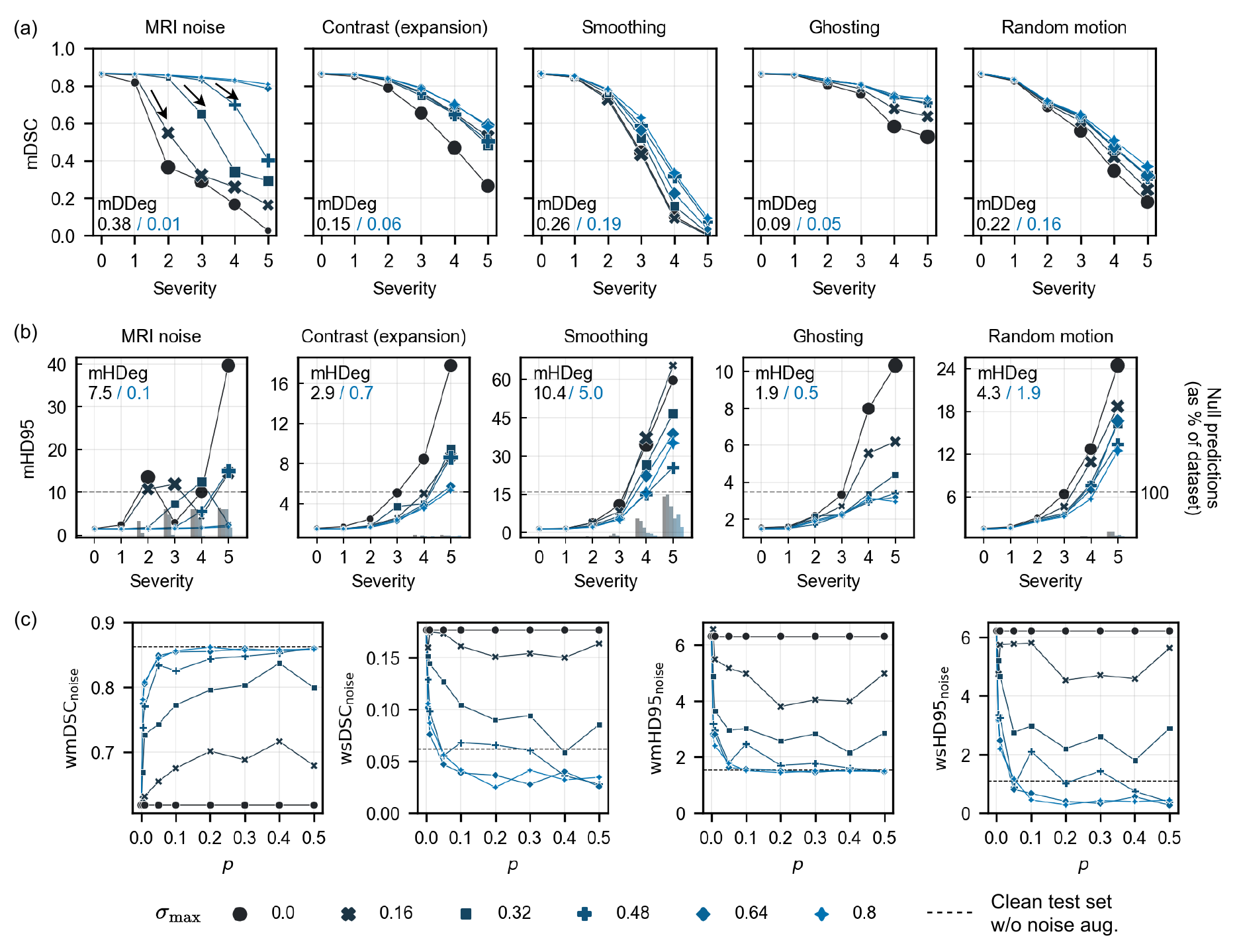}
\caption{Hippocampus segmentation robustness improves with noise augmentation during training. (a) $\mathrm{mDSC}$ and (b) $\mathrm{mHD95}$ improve substantially with increasing intensity range of the noise ($\sigma_{\mathrm{max}}$) applied to samples during training (shown here for augmentation probability $p = 0.1$). Degradation metrics are shown in each panel for no augmentation (black) and nonzero $\sigma_{\mathrm{max}}$ (blue). Marker size (area) is proportional to variance-based metrics: (a) $\mathrm{sDSC}$ and (b) $\mathrm{sHD95}$. (c) Weighted metrics as a function of augmentation probability $p$ for different noise intensity ranges ($\sigma_{\mathrm{max}}$) applied during training.}
\label{fig:augmentation_single_transform_sensitivity_curves}
\end{figure}

Results are visualized in \autoref{fig:augmentation_single_transform_sensitivity_curves} for models trained and evaluated on the hippocampus dataset. In most cases, given a large enough augmentation frequency $p$, models acquired robustness to noise at test time according to the level of the corruptions they had seen during training. We observed that test-time $\mathrm{mDSC}$ in the face of MRI noise drops off sharply after the severity level corresponding to the maximum intensity range of noise applied during training, $\sigma_{\mathrm{max}}$ (indicated by arrows in \autoref{fig:augmentation_single_transform_sensitivity_curves}a). With that said, models trained with $\sigma_{\mathrm{max}} \in \{0.64, 0.80\}$ exhibited nearly perfect robustness to MRI noise at test time, with $\mathrm{mDDeg}_{\mathrm{noise}}$ improving from 0.38 to 0.01 and $\mathrm{mHDeg}_{\mathrm{noise}}$ improving from 7.5 to 0.1, for no augmentation ($\sigma_{\mathrm{max}} = 0.0$) and $\sigma_{\mathrm{max}} = 0.80$, respectively. For the hippocampus segmentation task, models trained strictly with noise augmentation also improved in robustness with respect to other transforms at test time including gamma expansion, smoothing, ghosting, and random motion (\autoref{fig:augmentation_single_transform_sensitivity_curves}a-b). However, this trend was not consistent across tasks (see Supplementary Figure S3 and S4). Importantly,  for high transform intensity ranges (i.e., $\sigma_{\mathrm{max}} \in \{0.64, 0.80\}$) robustness gains typically saturated around probability $p = 0.1$, with increased frequency of augmentation during training resulting in little return by the way of improved robustness to noise (\autoref{fig:augmentation_single_transform_sensitivity_curves}c). For instance, by probability $p = 0.1$,  $\mathrm{wmDSC}_{\mathrm{noise}}$ had reached $\mathrm{mDSC}_{\mathrm{clean}}$ for models trained with no noise augmentation. Similarly, $\mathrm{wmHD95}_{\mathrm{noise}}$ reached $\mathrm{mHD95}_{\mathrm{clean}}$ by the same probability. While for variance-based metrics, by probability $p = 0.1$, $\mathrm{wsDSC}_{\mathrm{noise}}$ and $\mathrm{wsHD95}_{\mathrm{noise}}$ decreased beyond $\mathrm{sDSC}_{\mathrm{clean}}$ and $\mathrm{sHD95}_{\mathrm{clean}}$ for models trained with no noise augmentation, respectively (see \autoref{fig:augmentation_single_transform_sensitivity_curves}c).

\subsubsection{Augmentation with multiple transforms}
\label{sec:augmentation_multiple}

Next, we extended the experiments from Section \ref{sec:augmentation_single} to include all transforms studied in this work during augmentation. The augmentation scheme corresponding to this experiment is illustrated in \autoref{fig:augmentation_scheme}b. Instead of applying the same transform every time with probability $p$ (albeit with varying degrees of intensity), the probability $p$ indicates the frequency with which a transform will be applied \textit{at all}. Once the algorithm is triggered to apply a transform during training, one of the 11 transforms is randomly selected (with equal probability for each transform); then, parameters controlling the intensity of the transform are sampled and the transform is applied to the input image. Transform parameters were sampled uniformly for all transforms, ranging from the parameter value that corresponds to no transformation (e.g., $\sigma_g/\sigma_{\mathrm{img}} = 0$ for noise) and the parameter value corresponding to the maximum severity level in our benchmarking test sets. With the transform intensity range fixed (based on results from Section \ref{sec:augmentation_single}), the only parameter we varied in this experiment was the augmentation probability $p$. We studied whether larger probabilities $p$ would be required to attain the same degree of robustness on the single transform we studied in Section \ref{sec:augmentation_single}, given the fact that more distinct transforms are now being applied with the same overall augmentation frequency. In addition, we investigated the robustness improvement achieved towards all transforms by employing this combined augmentation strategy.

\begin{figure}[t!]
\centering
\includegraphics[width=\textwidth]{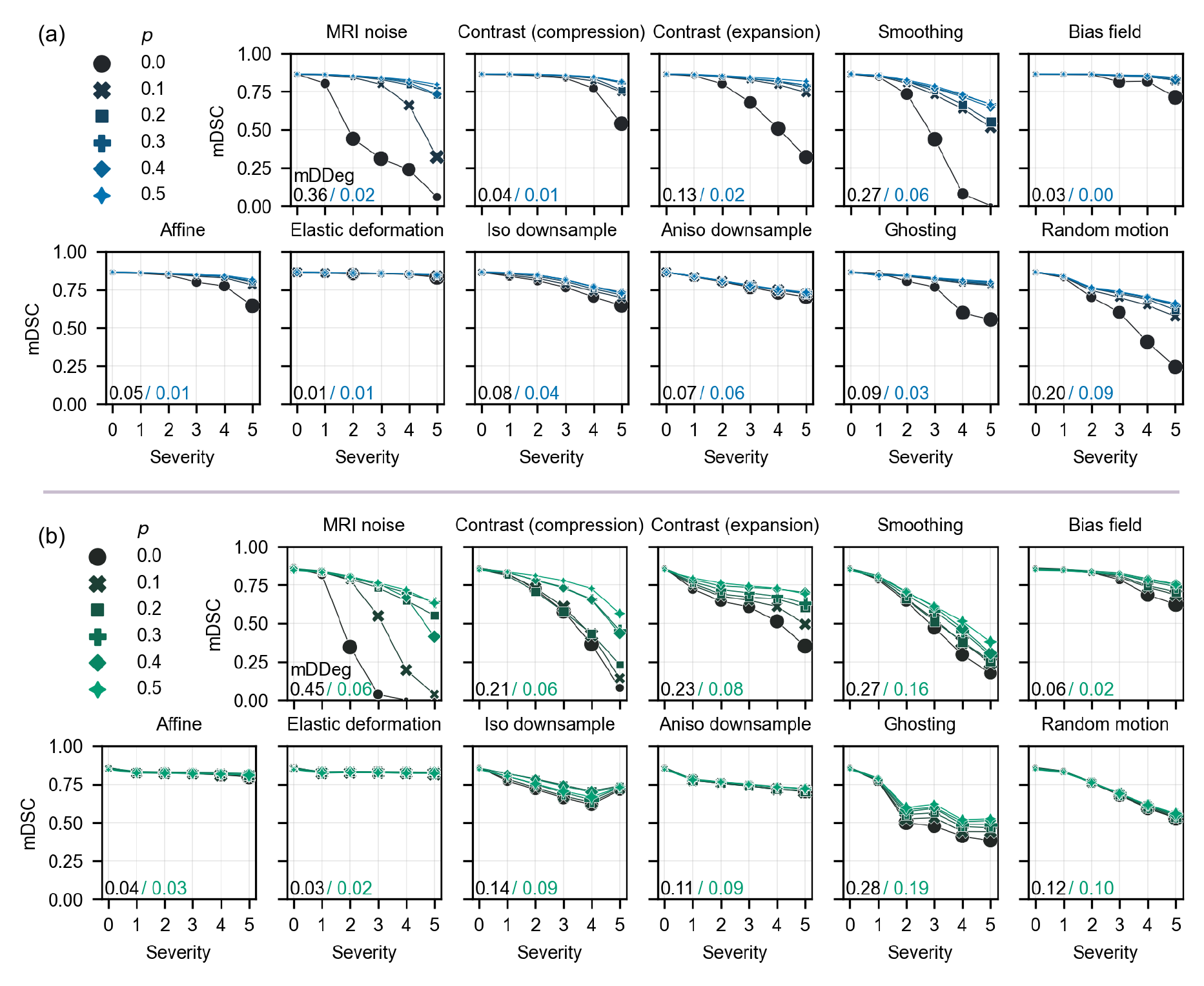}
\caption{Mean Dice similarity coefficient ($\mathrm{mDSC}$) sensitivity curves before and after including all transforms in data augmentation during training for (a) hippocampus and (b) WMH segmentation. Marker sizes (area) are proportional to standard deviation in DSC ($\mathrm{sDSC}$). Degradation metric values are shown for no augmentation (black) and nonzero $p$ (blue for hippocampus; green for WMH).}
\label{fig:augmentation_all_transforms_sensitivity_curves}
\end{figure}

For the hippocampus segmentation task, robustness to most transforms improved using the combined augmentation strategy (\autoref{fig:augmentation_all_transforms_sensitivity_curves}), although a higher augmentation probability $p$ was required to attain similar degrees of robustness improvement to what was observed using a single transform (Section \ref{sec:augmentation_single}). Specifically, robustness improvements saturated around probability $p = 0.4$. Transforms where the model failed to reach $\mathrm{mDDeg}_T \leq 0.03$ with $p = 0.5$ included smoothing, isotropic/anisotropic downsampling, and random motion. For the WMH segmentation task, robustness improvements were noticeably less pronounced than in the hippocampus task, although robustness to most transforms improved to some degree (\autoref{fig:augmentation_all_transforms_sensitivity_curves}). It is worth noting here that robustness to most transforms was lower to begin with for the WMH task relative to hippocampus segmentation (\autoref{fig:baseline_sensitivity_curves}). Weighted performance metrics, aggregated across all transforms, approached the performance on the clean test set with higher augmentation probability across all tasks (see Supplementary Figure S7) In some cases (e.g., for the variance-based metric $\mathrm{wsDSC}$ on hippocampus segmentation), clean performance improved substantially with larger $p$; while in others (e.g., variance-based $\mathrm{wsDSC}$ on WMH segmentation), the combined augmentation strategy did not produce a drastic improvement in performance, highlighting the need of exploring additional augmentation strategies or network architectures to attain further robustness for some tasks. Additional DSC sensitivity curves for the ventricle segmentation task and HD95 sensitivity curves for all tasks are shown in Supplementary Figure S5 and S6.

\subsection{Effect of encoder architecture on robustness: U-Net vs. Vision Transformer}
\label{sec:res:architectures}

\begin{figure}[t!]
\centering
\includegraphics[width=\textwidth]{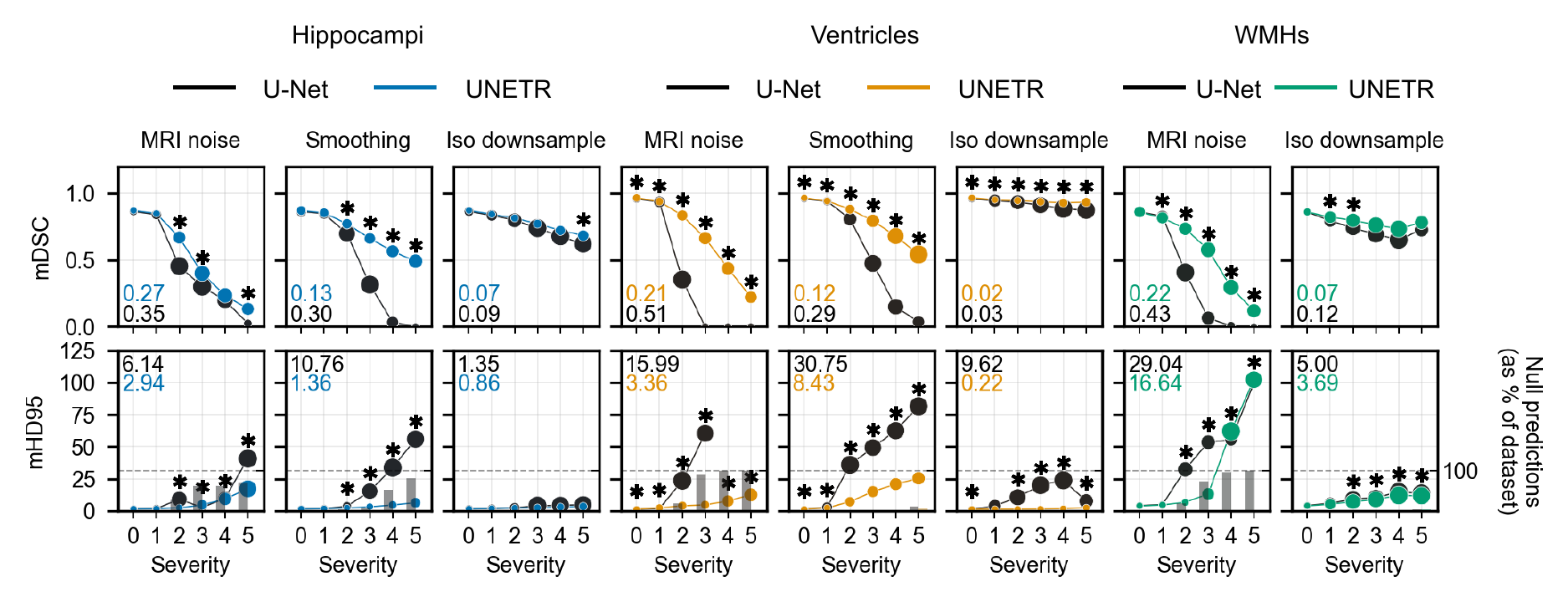}
\caption{Differences in robustness against select transforms across model architectures trained without transform-specific data augmentation, U-Net vs. UNEt TRansformer (UNETR), on the three tasks (Hippocampus, Ventricles and WMH); highlighting the significantly higher UNETR robustness on the presented transforms. Performance of the U-net on mean DSC (top row) and mean modified Hausdorff distance (HD95) is shown in black across the three tasks. UNETR performance is depicted in color based on the task (blue: hippocampi; orange: ventricles; green: WMHs). Asterisks represent a Wilcoxon signed-rank test p-value $< 0.01$ between models, with Bonferroni correction to account for multiple comparisons across severity levels. Numbers inside each axis corner correspond to mean-based $\mathrm{DSC}$ degradation ($\mathrm{mDDeg}$; top) and $\mathrm{HD95}$ degradation ($\mathrm{mHDeg}$; bottom) for both the UNETR (colored) and U-Net (black). Marker size (area) is proportional to $\mathrm{sDSC}$ (top) and $\mathrm{sHD95}$ (bottom).}
\label{fig:unet_vs_unetr_sensitivity_curves}
\end{figure}

In this experiment, we investigated the effects of encoder architecture on segmentation robustness, comparing the U-Net and UNet TRansformer (UNETR) described in Section \ref{sec:net_architectures}. To isolate the effect of network architecture, no data augmentation was applied during training with the exception of random flipping in each principal axis. Metrics on the clean test set as well as weighted metrics aggregated across all transforms are shown in \autoref{tab:unet_unetr_metrics} for all three tasks. Across tasks and transforms, there were several instances where the UNETR was significantly more robust than the U-Net. In most cases, robustness differences were significant for transforms that disrupt high-frequency spatial information (e.g., noise, blurring, downsampling). \autoref{fig:unet_vs_unetr_sensitivity_curves} shows sensitivity curves for tasks and transforms where the difference ($\Delta$) in mean- and variance-based robustness metrics ($\Delta\mathrm{mDDeg}_T$ and $\Delta\mathrm{vDDeg}_T$) between UNETR and U-Net were larger than $0.05$ and $0.04$, respectively. For some transforms, $\Delta\mathrm{mDDeg}$ was as large as 0.30. On smoothed data for the ventricle segmentation task, the UNETR was able to maintain $\mathrm{mDSC} > 0.50$ for all severity levels, whereas the U-Net dropped below this mark ($\mathrm{mDSC} < 0.50$) at severity level 3, reaching $\mathrm{mDSC} = 0.04$ by severity level 5. On isotropically downsampled data for the same task, the U-Net degraded to $\mathrm{mHD95} = 23.8$ mm by severity level 4 whereas the UNETR maintained $\mathrm{mHD95} < 2.3$ mm for all five severity levels. Differences in variance robustness were observed as well, notably with isotropic downsampling on anatomical segmentation tasks. Differences in robustness between networks were less pronounced on the WMH segmentation task, although some transforms (noise, isotropic downsampling) exhibited $\Delta\mathrm{mDDeg}_T > 0.05$.

\begin{table}[h!]
    \centering
    \caption{Performance on clean and transformed (summarized by weighted metrics) test sets for the U-Net and UNETR models. Null (empty) predictions are tabulated in the rightmost column as a percentage of all samples in all test datasets.}
    \vspace{2mm}
    \begin{threeparttable}
        \resizebox{\textwidth}{!}{
        \begin{tabular}{lllllllllll}
        \hline
              Task &     Model &   $\mathrm{mDSC}_{\mathrm{clean}}$ &  $\mathrm{wmDSC}$ &  $\mathrm{sDSC}_{\mathrm{clean}}$ &  $\mathrm{wsDSC}$ &  $\mathrm{mHD95}_{\mathrm{clean}}$ &  $\mathrm{wmHD95}$ &  $\mathrm{sHD95}_{\mathrm{clean}}$ &  $\mathrm{wsHD95}$ &  Null preds. (\%) \\
        \hline
        Hippocampi & U-Net &        0.86 &   0.79 &        0.05 &   0.08 &          1.5 &     3.0 &          0.8 &     2.3 &              7.4 \\
         &   UNETR &        \textbf{0.87} &   \textbf{0.81} &        \textbf{0.03} &   \textbf{0.06} &          \textbf{1.3} &     \textbf{2.3} &          \textbf{0.4} &     \textbf{1.9} &              \textbf{0.3} \\
         \hline
        Ventricles & U-Net &        0.96 &   0.88 &        0.02 &   0.06 &          1.2 &     5.5 &          0.4 &     4.9 &              7.8 \\
         &   UNETR &        \textbf{0.97} &   \textbf{0.91} &        0.02 &   \textbf{0.05} &          \textbf{1.1} &     \textbf{2.9} &          0.4 &     \textbf{2.9} &              \textbf{0.9} \\
         \hline
              WMHs & U-Net &        0.86 &   0.74 &        \textbf{0.08} &   \textbf{0.12} &          4.1 &    10.3 &          5.6 &     9.5 &              6.1 \\
                &   UNETR &        0.86 &   \textbf{0.76} &        0.09 &   0.13 &          \textbf{3.9} &     \textbf{9.3} &          \textbf{5.1} &     9.5 &              \textbf{1.7} \\
        \hline
        \end{tabular}}
        \label{tab:unet_unetr_metrics}
    \end{threeparttable}
\end{table}

\begin{figure}[t!]
\centering
\includegraphics[width=\textwidth]{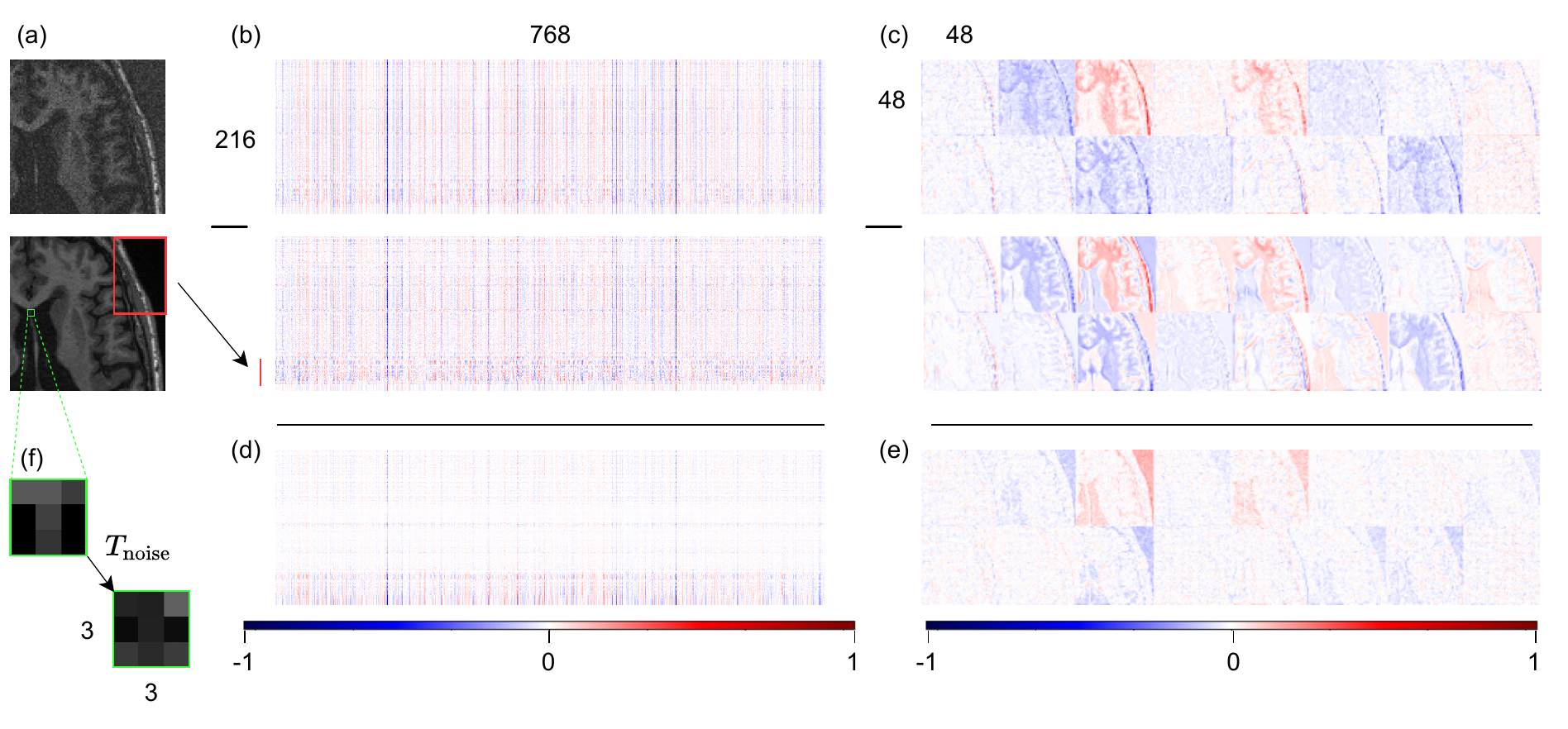}
\caption{Differences in encoder features between UNETR and U-Net models trained on ventricle segmentation after transforming data with noise corruption ($\sigma_g/\sigma_{\mathrm{img}} = 0.32$). (a) Central axial slices from identical $96 \times 96 \times 96$ mm\textsuperscript{3} patches corrupted with noise (top) and clean (bottom). (b) Mini patch embeddings (216 mini patches, length-768 encoding vectors) after the linear embedding layer of the UNETR ViT for the patches in (a). (c) Central axial slices of the 16 feature maps produced by the first convolutional layer of the U-Net. To compare across models, both sets of feature maps were scaled to the range $[-1, 1]$. (d)-(e) Element-wise differences between corrupted and clean encodings for (b) and (c), respectively. (f) A characteristic ``T'' shaped feature between lateral ventricles is indistinguishable at the local scale of a convolution kernel after corruption with noise.}
\label{fig:unet_vs_unetr_features}
\end{figure}

To further investigate the difference in robustness to the transforms in \autoref{fig:unet_vs_unetr_sensitivity_curves} between UNETR and U-Net models, we visualized early encoding features produced by both networks on transformed and clean data (\autoref{fig:unet_vs_unetr_features}). Images correspond to the same sample (clean vs. transformed with MRI noise at $\sigma_g/\sigma_{\mathrm{img}} = 0.32$) and encoding features correspond to models trained on the ventricle segmentation task. Qualitatively, we can see that the main difference between feature encodings for the UNETR on clean and transformed data are for the band corresponding to the background region that lies outside the skull. The rest of the mini patch encodings, which correspond to areas inside the skull, appear similar between clean and transformed data. Meanwhile, differences in the initial layer of feature maps for the U-Net model are highly dissimilar between clean and transformed data in multiple areas. While feature maps are dissimilar for areas outside the skull (similar to the UNETR encodings), they are also markedly different in the areas corresponding to the ventricles themselves (region/task of interest). As many of the feature maps appear to be concerned with distinguishing ventricle and background regions from brain tissue and structures, the U-Net appears to lose some of its ability to distinguish between these two categories in the case of noise. \autoref{fig:unet_vs_unetr_features}f demonstrates how a characteristic ``T'' shaped feature between the two ventricles on clean data can become indistinguishable at the local $3 \times 3 \times 3$ convolution-kernel level when data is corrupted by noise.

\section{Discussion}
\label{sec:discussion}

In this work, we take the first step in characterizing the sensitivity of modern DNNs to specific types of OOD data, corruptions, and artifacts in MRI. Specifically, we developed a methodology for benchmarking models based on robustness to image corruptions and transforms simulating out-of-distribution data in MRI. Beyond incorporating MRI-specific transforms and severity levels, we derived metrics that are applicable in two different scenarios: (1) where overall performance -- accounting for both clean and transformed test data -- is desired to determine the best model for deployment (e.g., a weighted mean $\mathrm{DSC}$ over transforms and severity levels, $\mathrm{wmDSC}$), and (2) where a measure of robustness agnostic to performance on the clean test set is desired (e.g., a mean-based $\mathrm{DSC}$ degradation relative to performance on a clean dataset, $\mathrm{mDDeg}$). We defined overlap- and distance-based analogs for both metric classes, considering the robustness of the mean and variance in model performance. We showed that it is important to consider each of these aspects when evaluating models for robustness to OOD data. Building on the work of \citet{hendrycks2019benchmarking} in computer vision, our methodology should act as a tool for researchers in medical imaging to evaluate and compare network architectures and design considerations based on robustness to the transforms studied here. In addition, it should enable researchers to generate their own task-specific benchmarking datasets, which can be used internally for high-throughput model evaluation or distributed to the medical imaging community for widespread model comparison. 

\subsection{Sensitivity to OOD data}
\label{sec:dsc:sensitivity}

We demonstrated that DNNs are highly sensitive to characteristic MRI noise, changes in contrast and intensity non-uniformity that lie outside the training distribution, reduction in image resolution, and motion artifacts. It is important to note that the variance in model prediction performance also increases with the increasing severity of exposure to OOD data. As DNNs become deployed across different studies and integrated into multi-site and multi-scanner image analysis workflows, it will become increasingly important to be aware of these models' sensitivities and weaknesses, benchmark models in terms of generalizability to ODD data, and use best practices in designing models for robustness. While some transforms have been shown to degrade model performance rapidly as the severity level increases, we observed others with a much smaller effect. Interestingly, affine transformations and elastic deformations, which are frequently employed in data augmentation schemes \citep{shorten2019survey}, had little effect on model performance. We considered affine transformations up to $d = 40$ mm and $\theta = 30^{\circ}$, and random displacements up to $d = 30$ mm at $7^3$ equally-spaced control points throughout the image volume for elastic deformations. Data augmentation serves more than just the purpose of increasing robustness to prospective OOD data; it creates more unique data samples to train on, helping to prevent memorization and overfitting to the training set. Nevertheless, we observed that a typical residual U-Net trained with minimal augmentation (random flipping only) is already quite robust to affine transformations and elastic deformations in a test setting.

\subsection{False positives induced by distribution shift}
\label{sec:dsc:false_positives}

While the most common failure mode for models faced with OOD data was in relation to false-negative voxels/predictions (i.e., an inability to identify voxels corresponding to the feature of interest), several classes of transforms were also shown to induce false-positive predictions (i.e., identifying feature-of-interest voxels where there are none). Most notably, motion artifacts such as ghosting could result in the repetition of features (such as the skull) that resemble the feature of interest (e.g., WMHs) if the repeated structures are positioned accordingly. Other classes of OOD data, such as changes in contrast and image resolution, were capable of inducing false positives in both the WMH and ventricle segmentation tasks. Such errors could be more challenging to detect than their false-negative counterparts in a clinical deployment setting focused on accurate volumetrics. While under-segmented volumes may present as outliers due to low counts of segmented voxels, predictions with false positives may fit within the expected segmented volume distribution, and would require visual inspection or quality control to be flagged for error. This is particularly important when deploying models on OOD data with distribution shifts that induce a large number of false positives (e.g., motion artifacts, reduction in image resolution, nonlinear changes in contrast).

\subsection{Effect of patch size and spatial context on robustness}
\label{sec:dsc:patch_size}

Many deep learning models in neuroimaging are trained on fixed-size image patches out of necessity. 3D image volumes and high-resolution 2D slices are often too large for the GPU's memory to facilitate a forward and backward pass when training a large network. While some modern GPUs can fit 3D image volumes the size of typical neuroimaging scans into memory, this usually limits the training batch size to one. Meanwhile, patch-based models can train on batches that are on the order of several to tens or hundreds of patches, depending on patch size and the dimensionality of the data. In deep CNNs, receptive fields can reach the full size of the image volume (e.g., $256 \times 256 \times 256$) by the end of the network; however, the receptive fields for patch-based models are limited by the size of the extracted patch.

We expected models with large patch sizes to show more robustness to transforms that disrupt local information and where long-range spatial context could help the model make accurate predictions at a given voxel. Our results confirmed this hypothesis: U-Nets using a patch size of $32 \times 32 \times 32$ mm\textsuperscript{3} (32P) demonstrated a markedly lower degree of robustness to \textit{blurring} transforms (e.g., smoothing, downsampling, random motion) in comparison to those using a larger patch size (e.g., 64P, 96P). These transforms all share in common that they disrupt high-frequency spatial information. These results suggest that a larger spatial context may be necessary to extract important spatial patterns and features when low-level information is lost. Meanwhile, our results also suggest that the dichotomy between whole-image and patch-based models may not be necessary for specific segmentation tasks. Patch-based models with a large enough patch size (64P, 96P) performed just as well or better than whole-image models on clean data and showed nearly identical degrees of robustness to OOD data. This result further highlights that given large patch sizes (relative to the feature-of-interest) the context of the entire image may not be necessary to make accurate predictions when OOD transforms or corruptions disrupt low-level features of the image. 

\subsection{Data augmentation improves robustness}
\label{sec:dsc:augmentation}

Simple data augmentation strategies during training generally improved segmentation robustness to OOD data across tasks and transforms. First, using a single transform during augmentation, we investigated the relationship between two critical parameters: the intensity range from which augmentation transform parameters were sampled from, and the probability $p$ of applying the transform to a given sample during training. We found that no matter how large $p$ was, models were only able to attain robustness at test time to the level of severity of the augmentation transform seen during training. For example, models will only maintain near-clean performance at test time on data corrupted with noise up to $\sigma_g/\sigma_{\mathrm{img}} = 0.32$ if $\sigma_g/\sigma_{\mathrm{img}} = 0.32$ was the most intense distortion seen during training. Therefore, it is important to be aware of the maximum distribution shift expected to be seen by a deployed model, and include augmentation transforms simulating the full range of distribution shift during training.

Next, we investigated whether inclusion of all the transforms in our benchmarking datasets during augmentation could eliminate significant performance degradation at test time on our benchmarking datasets. For hippocampus and ventricle segmentation tasks, overall mean-based $\mathrm{DSC}$ degradation (across all transforms) decreased from 0.12 to 0.03, and 0.13 to 0.03, respectively. Improvements on distance-based ($\mathrm{HD95}$) degradation metrics for the same tasks were of a similar magnitude (each greater than 75\% improvement). Meanwhile, robustness improvements on the WMH (lesion-based) segmentation task were smaller, with the overall mean-based $\mathrm{DSC}$ degradation decreasing from 0.18 to 0.08 (55\% improvement) and the overall mean-based $\mathrm{HD95}$ degradation decreasing from 9.0 to 3.3 (63\% improvement). These results suggest that additional augmentation strategies or network architectures may be needed to attain further robustness for some tasks. While \citet{zhang2020generalizing} demonstrated that their ``BigAug'' strategy improved segmentation performance on data from extrinsic sites on prostate and left-atrial MRI segmentation tasks, we explored the effect of data augmentation on robustness to individual transforms and corruption types, finding that on certain tasks, augmentation strategies combining several transforms (11 in our case) resulted in greater robustness at test time to particular transforms more than others. For example, on WMH segmentation, mean-based $\mathrm{DSC}$ degradation improvements were better for intensity-based transforms at test time (e.g., 0.45 to 0.06 and 0.21 to 0.06 on MRI noise and gamma compression, respectively) compared to motion artifacts (e.g., 0.28 to 0.19 and 0.12 to 0.10 for ghosting and random motion transforms, respectively). This demonstrates the importance of having a platform where model robustness can be compared against specific distribution shifts and corruption types, especially when anticipated distribution shift types in a real-world deployment setting are known.

\subsection{Encoder architecture: fully convolutional vs. transformer-based processing}
\label{sec:dsc:architectures}

After years of dominance from fully convolutional network (FCN) architectures such as the U-Net, Vision Transformer (ViT)-based models have recently emerged as a powerful alternative for image recognition \citep{dosovitskiy2020image} and segmentation \citep{hatamizadeh2021unetr} tasks. While \citet{hatamizadeh2021unetr} demonstrated favorable benchmarks on clean test data for the UNEt TRansformer (UNETR) over other FCN-based architectures on brain tumor and spleen segmentation tasks, the UNETR architecture has yet to be evaluated for its robustness to OOD data. There is a history of more complex architectures demonstrating little robustness improvement in computer vision tasks despite increases in clean test performance \citep{hendrycks2019benchmarking}. Nevertheless, given that one of the unique advantages of ViTs is their ability to model longer-range spatial relationships and dependencies through self-attention mechanisms, we hypothesized that the UNETR might incur more robustness to OOD shifts and corruptions that disrupt local information relative to FCN-based architectures such as the U-Net which operate on local receptive fields.

Our results demonstrated that the UNETR was significantly more robust, without any transform-specific data augmentation, than the U-Net implementation we employed across several transforms that disrupt low-level spatial information (e.g., noise, smoothing, downsampling). While part of this effect may have to do with the modeling of long-range spatial relationships in the image by ViTs, we also observed through visualization that features from the first encoder layers may be less disturbed by corruptions such as noise in ViTs as opposed to FCNs. We hypothesize that this effect may be due to the dimensionality of the dot products that form the basis of both convolution- and transformer-based operations. In CNNs, convolution kernels (filters) operate on small, local regions of an image (typically $3 \times 3 \times 3$ voxels; see \autoref{fig:unet_vs_unetr_features}). Noise added to the image can greatly affect the output of a dot product with only 27 elements. Meanwhile, the initial encoding layer of the ViT studied here performs dot products on large vectors of length 4096 where there are more elements for noise to be averaged out in a weighted sum. Subsequent self-attention and multilayer perceptron operations in the ViT continue to operate on vectors in a high-dimensional embedding space (vectors $v_i \in \mathbb{R}^{768}$) and may further help lessen the effect of noise corruption. These effects warrant further investigations in other segmentation tasks, modeling additional domain shifts and transforms. Meanwhile, the improvements in robustness demonstrated in this work attributed to the UNETR architecture could influence its adoption in more image segmentation pipelines, especially when robustness to OOD data is a key requirement.

While the UNETR (Vision Transformer) and U-Net showed similar degradation in mean performance when faced with isotropically downsampled data on anatomical segmentation tasks, the UNETR maintained significantly lower variance in both overlap- and distance-based metrics across test sets. Differences in robustness to isotropically downsampled data on the ventricle segmentation task between the UNETR and U-Net were more pronounced when the Hausdorff distance was considered instead of the Dice similarity coefficient. These results demonstrate the utility of having a suite of metrics that account for diverse characteristics of segmentation predictions and model performance.

\subsection{Limitations and future work}
\label{sec:dsc:limitations}

One of the main limitations of this work is regarding the degree to which the transforms and severity levels included in our methodology cover the range of distribution shifts observed in practice from different sites and scanners. While we included a reasonably large set of transforms that we deemed most relevant in practice from our experience, there could be other MRI transforms and corruptions that are relevant and under-represented in our list. Certain distribution shifts, such as changes in contrast, may be modeled better by other nonlinear transformations besides gamma correction. Our platform, however, is flexible to the implementation or addition of other transforms and corruptions, as well as changes in ranges of severity levels. Ideally, severity levels would correspond to the common range of distribution shifts observed in practice across multiple imaging sites. This modeling is difficult to achieve given the challenges in translating observed variances to parametrically-modeled transforms; however, further studies quantifying the characteristics of distribution shifts and corruptions in MRI may enable more comprehensive modeling. Such studies may also enable the weighting function across severity levels to be adjusted to model the distribution of observed variances across relevant sites. We intended the transforms and severity levels here to approximate shifts seen in practice and allow for comparison of relevant robustness considerations and design choices when developing models across different studies and for multi-site and multi-scanner deployment.

We intend for this platform to be used by the research community to generate benchmarking datasets for other tasks and datasets besides those used in this work. We designed the transforms and severity levels in a way such that they are broadly applicable across different anatomical regions and MR pulse sequences. Future work might involve extending this methodology and platform to other imaging modalities including CT and PET, including modality-specific distribution shifts and corruptions. Lastly, while we covered common augmentation strategies in this work, there are a number of more complex strategies involving mixing of transforms and complex sampling strategies that could be explored for their potential to increase robustness to OOD data in MRI. The prospect of using new network architectures for segmentation such as the Vision Transformer brings about a host of new questions regarding which architectural choices contribute to OOD robustness. Future work will focus on benchmarking more architectures on our datasets, and performing ablation studies to help elucidate important architectural features that result in robustness improvements.

\section{Conclusions}
\label{sec:conclusions}

In this work, we present ROOD-MRI: a novel benchmarking platform for evaluating the robustness of segmentation DNNs to OOD data, corruptions and artifacts in MRI. This platform enables comparing across models and generating new benchmarking datasets within the community, which can in turn be used to study architectural and model-design considerations that result in improved robustness. We demonstrated that modern DNNs are highly susceptible to transforms modeling distribution shifts in signal-to-noise ratio, contrast and intensity non-uniformity, image resolution, and presence of motion artifacts on segmentation tasks including hippocampus, ventricle, and WMH segmentation. Not only does the mean model performance decrease with increasing severity level of these transforms, but the variance substantially increased in transformed test sets on the three tasks. We demonstrated that U-Nets trained on small patch sizes are less robust to blurring transforms (e.g., smoothing, downsampling, motion artifacts) than similar models trained on larger patch sizes or the whole image, suggesting that spatial context may play a role in robustness against transforms that obscure low-level spatial information. Notably, we demonstrated that while common data augmentation strategies during training have the capability to improve model robustness without having to modify the network architecture, these robustness improvements were transform-specific and CNNs with augmented data were still highly susceptible to certain distribution shifts in the lesion-based task (WMH segmentation). Finally, by comparing the UNETR and U-Net architectures on our benchmarking datasets, we found that transformer-based networks may be more robust to distribution shifts and corruptions that disrupt low-level spatial information (e.g., noise, smoothing, and downsampling). The transforms, metric implementations, and modules used for generating benchmarking datasets and comparing across models are made publicly available for the research community.

\section*{Declaration of competing interest}

The authors declare that they have no known competing financial interests or personal relationships that could have appeared to influence the work reported in this paper.

\section*{CRediT authorship contribution statement}

\textbf{Lyndon Boone:} conceptualization, methodology, software, validation, formal analysis, investigation, writing -- original draft, writing -- review \& editing, visualization, project administration. \textbf{Mahdi Biparva:} conceptualization, methodology, software, writing -- review \& editing. \textbf{Parisa Mojiri Forooshani:} conceptualization, methodology, validation, writing -- review \& editing. \textbf{Joel Ramirez:} resources, data curation, writing -- review \& editing. \textbf{Mario Masellis:} resources, data curation, writing -- review \& editing. \textbf{Robert Bartha:} resources, data curation, writing -- review \& editing. \textbf{Sean Symons:} resources, data curation, writing -- review \& editing. \textbf{Stephen Strother:} resources, data curation, writing -- review \& editing. \textbf{Sandra E. Black:} resources, data curation, writing -- review \& editing, supervision. \textbf{Chris Heyn:} writing -- review \& editing, supervision. \textbf{Anne Martel:} conceptualization, writing -- review \& editing, supervision. \textbf{Richard H. Swartz:} conceptualization, writing -- review \& editing, supervision. \textbf{Maged Goubran:} conceptualization, methodology, formal analysis, resources, writing -- original draft, writing -- review \& editing, visualization, supervision, project administration, funding acquisition.

\section*{Acknowledgments}

This study was funded by the Natural Sciences and Engineering Research Council (NSERC) Discovery Grant \#RGPIN-2021-03728, the L.C Campbell Foundation and the SEB Centre for Brain Resilience and Recovery. LB is supported by the Alexander Graham Bell NSERC CGS-M scholarship. MG is supported by the Gerald Heffernan foundation and the Donald Stuss Young Investigator innovation award. RHS is supported by a Heart and Stroke Clinician-Scientist Phase II Award. This research was enabled in part by support provided by WestGrid (\url{www.westgrid.ca}) and Compute Canada (\url{www.computecanada.ca}). This research was conducted with the support of the Ontario Brain Institute, an independent non-profit corporation, funded partially by the Ontario government. The opinions, results and conclusions are those of the authors and no endorsement by the Ontario Brain Institute is intended or should be inferred. Matching funds were provided by participant hospital and research foundations, including the Baycrest Foundation, Bruyere Research Institute, Centre for Addiction and Mental Health Foundation, London Health Sciences Foundation, McMaster University Faculty of Health Sciences, Ottawa Brain and Mind Research Institute, Queen’s University Faculty of Health Sciences, St. Michael's Hospital, Sunnybrook Health Sciences Centre Foundation, the Thunder Bay Regional Health Sciences Centre, University Health Network, the University of Ottawa Faculty of Medicine, and the Windsor/Essex County ALS Association. The Temerty Family Foundation provided the major infrastructure matching funds. We are grateful for the support of the Medical Imaging Trial Network of Canada (MITNEC) Grant \#NCT02330510, and the following site Principal investigators: Christian Bocti, Michael Borrie, Howard Chertkow, Richard Frayne, Robin Hsiung, Robert Laforce, Jr., Michael D. Noseworthy, Frank S. Prato, Demetrios J. Sahlas, Eric E. Smith, Vesna Sossi, Alex Thiel, Jean-Paul Soucy, and Jean-Claude Tardif. We are also grateful for the support of the Canadian Atherosclerosis Imaging Network (CAIN) (\url{http://www.canadianimagingnetwork.org/}), and the following investigators: Alan Moody, Therese Heinonen, Rob Beanlands, David Spence, Philippe L’Allier, Brian Rutt, Aaron Fenster, Matthias Friedrich, Ben Chow, and Richard Frayne. 

\bibliography{ms}

\appendix

\section{Transform formulations}
\label{app:transforms}

Across transform formulations, let $\mathbf{x} \in \mathcal{X}$ represent a particular 3D voxel coordinate, where $\mathcal{X} = \{(1, \ldots, H) \times (1, \ldots, W) \times (1, \ldots, D)\}$ is the space of all such voxel coordinates in an image volume. Where transformation equations are provided, let $I \in \mathcal{I}$ represent the clean input image volume, and let $S \in \mathcal{I}$ represent the transformed image volume, where $\mathcal{I} = \{I\;|\;I: \mathcal{X} \rightarrow \mathbb{R}\}$ is the space of all image volumes mapping three-dimensional coordinates (e.g., $\mathbf{x} \in \mathcal{X}$) to scalar voxel-intensity values. Accordingly, $I(\mathbf{x}) \in \mathbb{R}$ and $S(\mathbf{x}) \in \mathbb{R}$ represent the voxel intensity at coordinate $\mathbf{x} \in \mathcal{X}$ for the input image and transformed image, respectively.

\paragraph{MRI noise}

Voxel intensities in noisy magnitude MR images follow the Rician distribution \citep{gudbjartsson1995rician, rice1944mathematical, cardenas2008noise}. This phenomenon is the result of magnitude image formation from complex data with independent and identically distributed Gaussian noise in each channel. We modeled MRI noise by computing the transformed image as follows \citep{garyfallidis2014dipy}:
\begin{equation}
    S(\mathbf{x}) = |I(\mathbf{x}) + N_1(\mathbf{x}) + iN_2(\mathbf{x})|,
    \label{eq:noise}
\end{equation}
where $N_1(\mathbf{x}), N_2(\mathbf{x}) \sim \mathcal{N}(0, \sigma_g^2)$ are elements from two volumes $N_1$ and $N_2$ with the same shape as $I$ consisting of Gaussian noise with zero mean and variance $\sigma_g^2$. For each severity level, $\sigma_g/\sigma_{\mathrm{img}}$ was fixed to a constant, where $\sigma_{\mathrm{img}}$ is the standard deviation of the input image intensity histogram. This way, severity levels are relatively agnostic to the range and distribution of voxel intensities.

\paragraph{Contrast}

Changes in contrast were modeled using the gamma correction technique \citep{gonzalez2006digital}. While gamma correction may be used as a post-processing technique in medical imaging to alter contrast between features of interest in an image, we employed it as a means to model nonlinear shifts in contrast between images from scanners and acquisition protocols. Typically, gamma correction is applied to input images with intensities scaled to the range $[0, 1]$. As such, let $I_{\mathrm{min}} = \min_{\mathbf{x} \in \mathcal{X}}I(\mathbf{x})$ be the minimum voxel intensity in the input image $I$, and let $\Delta_I = \max_{\mathbf{x} \in \mathcal{X}} I(\mathbf{x}) - I_{\mathrm{min}}$ be the range of voxel intensities in $I$. The transformed image was computed as
\begin{equation}
    S(\mathbf{x}) = \bigg(\frac{I(\mathbf{x}) - I_{\mathrm{min}}}{\Delta_I}\bigg)^{\gamma} \cdot \Delta_I + I_{\mathrm{min}},
    \label{eq:gamma_correction}
\end{equation}
where $\gamma > 0$ parameterizes the transform. In this work, we treat $\gamma < 1$ and $\gamma > 1$ as distinct transforms, referring to them as \textit{gamma compression} and \textit{gamma expansion}, respectively.

\paragraph{Smoothing}

Structural MR images are often smoothed to increase SNR, initiating a tradeoff between noise reduction and detailed/high-frequency feature preservation \citep{mohan2014survey}. In this work, we also use smoothing as a proxy for images with intrinsically low spatial resolution. Smoothing was applied using a simple Gaussian filter:
\begin{equation}
    S(\mathbf{x}) = G(\mathbf{x}; \sigma) \ast I(\mathbf{x}),
    \label{eq:smoothing}
\end{equation}
where $G$ represents a Gaussian kernel parameterized by standard deviation $\sigma$, and $\ast$ represents 3D convolution. Gaussian kernels were truncated at $4\sigma$.

\paragraph{Bias field}

Bias field (BF), or intensity non-uniformity, is a common artifact in MRI characterized by smooth, low-spatial-frequency intensity variation across an image \citep{sled1998understanding}. We use the TorchIO implementation \citep{perez2021torchio}, which models BF as a third-order, multiplicative polynomial field, after \citet{van1999automated} and \citet{sudre2017longitudinal}:
\begin{equation}
    S(\mathbf{x}) = e^{B(\mathbf{x})} \cdot I(\mathbf{x}),
    \label{eq:biasfield}
\end{equation}
where $B$ represents a volume with the same shape as $I$ whose values correspond to a third-order polynomial function over the domain $[-1, 1]$ in each spatial axis. We randomly vary the bias field for each test image by sampling polynomial coefficients uniformly from $(-b, b)$, where $b$ parameterizes the severity of the transform.

\paragraph{Affine transformations}

Affine transformations are ubiquitous in MR and CT imaging. They are used to define the relation between voxel and scanner coordinate systems, and are commonly used as a default image registration transform in analysis pipelines. We used 3D affine transformations to simulate data that may not have been registered to the same space or orientation as the training data. We considered rigid rotations and translations only, as these can easily arise due to slight positioning differences between patients in the scanner (e.g., head tilt). The transformed image was computed as
\begin{equation}
    S(\mathbf{x}) = I(M\mathbf{x} + \mathbf{d}),
    \label{eq:affine}
\end{equation}
where $M \in \mathbb{R}^{3 \times 3}$ is a 3D rotation matrix and $\mathbf{d} \in \mathbb{R}^3$ is a translation vector. Interpolation is implicit in \autoref{eq:affine}, used to sample between points in the input image. Unless stated otherwise, we use linear interpolation. For each test image in a given benchmarking dataset, a random affine transformation was applied by sampling Euler angles uniformly from $(-\theta, \theta)$ and sampling elements of $\mathbf{d}$ uniformly from $(-d, d)$. $\theta$ and $d$ parameterize the severity of the transform.

\paragraph{Elastic deformation}

Elastic deformation (ED) was used to simulate the high heterogeneity in neuroanatomy across the population and different ages, as well as brain or skull deformations caused by severe trauma, mass effect, or midline shift. We use the TorchIO implementation for ED \citep{perez2021torchio} which assigns a random displacement to each point from a coarse grid in and around the image, interpolating between points using cubic B-splines to obtain the output image. We kept the number of control points along each image axis constant at seven, while varying the range in which grid point displacements are sampled from to alter the severity. Displacements for each point within the grid were sampled uniformly from the range $(-d, d)$. The parameter $d$ controls the severity of the transform.

\paragraph{Downsampling}

Spatial resolution in MRI is controlled by several factors, including the number of frequency- and phase-encoding steps in an acquisition sequence, the field of view (FOV), and the slice thickness. Depending on the scanner, sequence, and acquisition parameters, each may vary significantly between sites and acquisition sessions. It is also common for one spatial axis to be sampled at a lower spatial resolution than the others. To address this, we apply two distinct transforms, which we refer to as \textit{isotropic} and \textit{anisotropic} downsampling (ID and AD, respectively). Each transform uses linear interpolation to downsample the image by a factor $F$, followed by upsampling back to the original image size. This procedure effectively reduces the inherent resolution of the image while maintaining the same voxel size and shape for consistency during further processing operations. For anisotropic downsampling, only one image axis is downsampled. The downsampled axis was randomly selected each time the transform was applied to an image volume when generating benchmarking datasets.

\paragraph{Ghosting}

Periodic motion (e.g., due to cardiac or respiratory cycles) can result in phase differences between readouts in k-space during acquisition \citep{zhuo2006mr}. These often appear as ghosting artifacts: faint, misaligned repetitions of structures in the image along the phase-encoding axis. Additionally, certain types of imaging sequences, such as echo planar imaging (EPI) involve fast switching of the readout direction (gradient switching), which can similarly lead to phase differences between readouts, ghosting artifacts, and eddy-current distortions. We use the TorchIO implementation for modeling ghosting \citep{perez2021torchio}, which simulates $N$ distinct ghosts by zeroing every $N$\textsuperscript{th} plane in k-space in the phase-encoding axis (preserving the center of k-space to avoid particularly severe artifacts). We controlled the severity of the transform by altering the number of ghosts in the image. The ghosting axis was randomly sampled from the left-right and anterior-posterior axes, as these are the most common phase-encoding axes in neuroimaging.

\paragraph{Random motion}

Similarly, random motion (RM) due to patient movement in the scanner can lead to discontinuities in the k-space trajectory. This type of movement causes less coherent motion artifacts than ghosting, which appear as a blurring of the image. We use the model by \citet{shaw2020k} for simulating motion artifacts in a clean image. Their algorithm involves splicing together $N$ segments of k-space formed by the inverse Fourier transform of $N$ affine-transformed copies of the clean image at different time points during the acquisition trajectory. For each simulated movement, Euler angles and elements of the displacement vector were sampled uniformly from $(-\theta, \theta)$ and $(-d, d)$, respectively. The parameters $N$, $\theta$, and $d$ are used to control the severity of the transform.

\end{document}